\documentclass[%
 reprint,
 amsmath,amssymb,
 aip,jcp,
floatfix,dvipsnames
]{revtex4-1}

\usepackage{graphicx}
\usepackage{dcolumn}
\usepackage{bm}
\usepackage{marginnote}

\usepackage{hyperref,color,pgfplots,siunitx,multirow,braket}
\usetikzlibrary{plotmarks}
\pgfplotsset{compat=1.13} 

\DeclareSIUnit\wn{\raiseto{-1}\cm}

\begin{document}

\title{Vibrational coherences in half-broadband 2D electronic spectroscopy: spectral filtering to identify excited state displacements}

\author{Dale Green}
\email{Dale.Green@uea.ac.uk}
\affiliation{Physics, Faculty of Science, University of East Anglia, Norwich NR4 7TJ, UK}

\author{Giovanni Bressan}
\affiliation{School of Chemistry, University of East Anglia, Norwich, NR4 7TJ, UK}

\author{Ismael A. Heisler}
\affiliation{Instituto de Física, Universidade Federal do Rio Grande do Sul, 91509-900 Porto Alegre, RS, Brazil.}

\author{Stephen R. Meech}
\affiliation{School of Chemistry, University of East Anglia, Norwich, NR4 7TJ, UK}

\author{Garth A. Jones}
\email{garth.jones@uea.ac.uk}
\affiliation{School of Chemistry, University of East Anglia, Norwich, NR4 7TJ, UK}

\begin{abstract}

    Vibrational coherences in ultrafast pump-probe (PP) and 2D electronic spectroscopy (2DES) provide insight into the excited state dynamics of molecules. Femtosecond coherence spectra (FCS) and 2D beat maps yield information about displacements of excited state surfaces for key vibrational modes. Half-broadband 2DES (HB2DES) uses a PP configuration with a white light continuum probe to extend the detection range and resolve vibrational coherences in the excited state absorption (ESA). However, interpretation of these spectra is difficult as they are strongly dependent on the spectrum of the pump laser and the relative displacement of the excited states along the vibrational coordinates. We demonstrate the impact of these convoluting factors for a model based upon cresyl violet. Careful consideration of the position of the pump spectrum can be a powerful tool in resolving the ESA coherences to gain insights into excited state displacements. The paper also highlights the need for caution in considering the spectral window of the pulse when interpreting these spectra.
    
\end{abstract}

\maketitle
\section{Introduction}

    Oscillatory features in ultrafast nonlinear spectra correspond to electronic, vibrational or vibronic coherences: coherently excited superpositions of states, called wavepackets, which inform about the structure and dynamics of  molecules.\cite{Scholes2017,Dean2017,Jumper2018}  Electronic coherences identify delocalised exciton states between coupled chromophores and can be used to determine electronic coupling strengths.\cite{Chenu2015,Mancal2020AQuantum} Mixing with the vibrational modes of the chromophores produces vibronic coherences, which are of great interest as they may be linked to enhanced rates of energy and electron transfer in photosynthetic systems and photovoltaic materials.\cite{Fassioli2014PhotosyntheticCoherence, Romero2014,Fuller2014VibronicPhotosynthesis,Ferretti2014TheSpectroscopy,Novoderezhkin2015,Dean2016,Arsenault2020VibronicII} In the absence of electronic coupling between chromophores, vibrational coherences correspond to the creation of vibrational wavepackets on the ground and excited electronic state potential energy surfaces (PES), where the strength of the vibrational coupling reveals the displacement of the excited state minimum along each mode coordinate, with respect to the ground state PES.\cite{Dean2015BroadbandBlue,Zhu2020ElectronicSpectroscopy} 

    In pump-probe spectroscopy (PP), the decay of vibrational wavepackets created by interaction with the pump pulse is detected by interaction with a probe pulse.\cite{Zewail2000Femtochemistry:Bond,Berera2009UltrafastSystems,Liebel2015a} Employing a spectrally broad white light continuum probe extends the range of excited state absorption, providing greater structural information on higher electronic excited states. These experiments are well understood using the semiclassical wavepacket theory introduced by Heller,\cite{HellerEricJ.1981TheSpectroscopy, Pollard1990WaveExperiments,Pollard1992TheoryBacteriorhodopsin} with a strong dependence on the frequency and phase of the pump pulse. Chirped pulses can selectively promote ground vs.\ excited state wavepackets, highlighting the importance of properly characterising the excitation pulse.\cite{Bardeen1998FemtosecondBacteriorhodopsin,Malkmus2005Chirp1,Horikoshi2007RapidPulses,Wand2010ChirpPerspective} Vibrational coherences in PP are isolated as the residuals of a global fit over delay time, removing non-oscillatory pathways. Fourier transform of the residuals then produces a femtosecond coherence spectrum (FCS), which reveals information about PESs such as the amount of anharmonicity.\cite{Turner2020BasisSpectra,Arpin2021SignaturesSpectra,Barclay2022CharacterizingSpectra} These spectra have nodes indicative of ground vs.\ excited state wavepackets, understood via the wavepacket formalism.\cite{Vos1993VisualizationSpectroscopy,McClure2014CoherentState, C.Arpin2015SpectroscopicOscillations, Rafiq2015ObservingReaction, Bressan2022PopulationNanorings} 

    2D electronic spectroscopy (2DES) involves the same third-order light-matter interaction but provides greater detail via time-ordering of the pump interactions to resolve the excitation (pump) axis.\cite{Biswas2022CoherentSpectroscopies} Fourier analysis of the oscillatory coherences thus yields a 2D beat map for each mode.\cite{Butkus2012a,Seibt2013,Seibt2013b,Egorova2014Self-analysisSignals,Sahu2023High-sensitivityScan} These maps have been useful for distinguishing between electronic, vibrational and vibronic coherences,\cite{Christensson2011,Turner2012a,Butkus2013,Butkus2014,DeA.Camargo2017,Butkus2017DiscriminationNanoring} and are complementary to FCS.\cite{Zhu2020ElectronicSpectroscopy} There has been significant development of 2DES in recent years, where the pump-probe, partially collinear, setup combines a white light continuum probe with a narrower pump spectrum in half-broadband 2DES (HB2DES).\cite{Tekavec2009Two-dimensionalProbe,Bressan2023Half-broadbandReduction} More readily available than full broadband, which has an extended range for both the pump and probe spectra,\cite{Kearns2017BroadbandKHz,Mewes2021BroadbandDyes,Carbery2019ResolvingSpectroscopy,Timmer2024FullResolution} HB2DES provides greater access and resolution to vibrational coherences in the excited state absorption (ESA) region, allowing access to new information about the displacement of higher excited states via the excited state wavepacket. But 2D beat maps may be misleading due to an intrinsic filtering of transitions which removes peaks as a result of the finite width of the pump spectrum,\cite{DeA.Camargo2017, Green2018SpectralModel} an effect which is exacerbated by the involvement of multiple excited states, making ESA coherences challenging to resolve and interpret.\cite{Bressan2024Two-DimensionalDisplacements}

    Here, we demonstrate the subtlety of FCS and 2D beat map analysis by exploring the effects of increasing S$_\text{n}$$\leftarrow$S$_1$ displacement and filtering due to the finite width of the pump spectrum relative to the steady state absorption band for a simple three state displaced harmonic oscillator (DHO) model based on the well-known dye molecule cresyl violet. We note that cresyl violet is often used as a molecule for benchmarking new ultrafast techniques.\cite{Turner2011, Carbery2019ResolvingSpectroscopy}

\section{Theory}

    We extend the DHO model to account for three electronic singlet states: the ground state S$_0$, $\ket{g}$, first excited state S$_1$, $\ket{e}$, and a higher lying excited state S$_\text{n}$, $\ket{f}$. The diabatic system Hamiltonian is,
    \begin{equation}
        H_S = \sum_i\ket{i}\!\bra{i}\otimes(E_i + h_i),
        \label{eq:H_S}
    \end{equation}
    with electronic energies $E_i$ and nuclear contributions,
    \begin{equation} \label{eq:hi}
        h_i=\hbar\omega_0\left(b^\dag b+\frac{1}{2}-\frac{\Delta_{ig}}{\sqrt{2}} (b^\dag+b)+\frac{1}{2} (\Delta_{ig})^2\right),
    \end{equation}
    where $i=\{g,e,f\}$. The three electronic states are coupled to a single harmonic vibrational mode with frequency $\omega_0$, which is assumed the same for all electronic states, and has associated vibrational creation, $b^\dagger$, and annihilation, $b$, operators. Coupling between the electronic and vibrational states results in displacement between the ground and excited state minima of $\Delta_{ig}$ along the dimensionless mode coordinate $Q = (b+b^\dagger)/\sqrt{2}$, where $\Delta_{gg}=0$.
    
    The diabatic electronic dipole moment operator,
    \begin{equation}
        \hat\mu_\text{el} = \Vec{\mu}_{eg}(\ket{e}\!\bra{g} + \ket{g}\!\bra{e}) + \Vec{\mu}_{fe}(\ket{f}\!\bra{e} + \ket{e}\!\bra{f}),
    \end{equation}
    accounts for allowed electronic transitions S$_1$$\leftarrow$S$_0$ and S$_\text{n}$$\leftarrow$S$_1$. For simplicity, the two transition dipole moments are assumed to be collinear and of equal magnitude, $\Vec{\mu}_{eg}=\Vec{\mu}_{fe}$. The total dipole moment operator $\hat{\mu}= \hat\mu_\text{el} \otimes I_\text{vib}$ where $I_\text{vib} = \sum_\nu\ket{\nu}\!\bra{\nu}$ is the identity operator for the vibrational mode. All system operators are transformed into the adiabatic basis via a unitary transformation which corresponds to diagonalisation of the system Hamiltonian.

    The system interacts with an environment modelled as an ensemble of harmonic oscillators separated into $N$ baths with overdamped spectral densities, approximated with the Lorentz-Drude form,
    \begin{equation} \label{eq:J_tot}
        J(\omega)=\sum_{n=1}^N 2\eta_n\frac{\omega\Lambda_n}{\omega^2+\Lambda_n^2},
    \end{equation}
    where $\eta_n$ is the reorganisation energy (coupling strength) of bath $n$ and $\Lambda_n$ is the dissipation rate associated with an exponential return to equilibrium of the system-bath correlation function. The inverse temperature, $\beta=(k_\text{B}T)^{-1}$, is assumed the same for all baths with $T=\SI{298}{K}$.
       
    To fully account for non-Markovian effects which determine the spectral broadening, a hierarchy of equations of motion is derived for the auxiliary density operators (ADOs), $\rho_\mathbf{j}$, as,\cite{Dijkstra2017,Green2019QuantifyingLineshape}
    \begin{eqnarray} \label{eq:Over_HEOMRate}
       \dot\rho_\mathbf{j}(t) &=& -\left(\frac{i}{\hbar}\mathcal{L} + \sum_{n=1}^N\sum_{k=0}^{M}{{j}_{nk}\nu_{nk}} \right)\rho_\mathbf{j}(t) \nonumber\\ 
       &\hspace{1em}& -i\sum_{n=1}^N\sum_{k=0}^{M}{{B}_n^\times\rho_{{j}_{nk}^+}(t)} \nonumber\\ &\hspace{1em}& -i\sum_{n=1}^N\sum_{k=0}^{M}{j}_{nk}\left(c_{nk}{B}_n\rho_{{j}_{nk}^-}(t)-c^\ast_{nk}\rho_{{j}_{nk}^-}(t){B}_n \right) \nonumber \\
        &\hspace{1em}& -\sum_{n=1}^N\left(\frac{2\eta_n}{\hbar\beta\Lambda_n}-\eta_n\cot\left(\frac{\hbar\beta\Lambda_n}{2}\right) \right. \nonumber \\
        &\hspace{1em}& \qquad \qquad \left. -\sum_{k=1}^{M}{\frac{c_{nk}}{\nu_{nk}}} \right){B}_n^\times {B}_n^\times\rho_\mathbf{j}(t), \hspace{2em}
    \end{eqnarray}
    where $\mathcal{L}\rho = {H}_\text{S}^{\prime\times}\rho = [{H}^\prime_\text{S},\rho]$, with the renormalised system Hamiltonian,\cite{Dijkstra2015,Tanimura2020NumericallyHEOM}
    \begin{equation} \label{eq:Ham_renorm}
    {H}_\text{S}^\prime={H}_\text{S}+\sum_n^N\eta_n{B}_n^2.
    \end{equation} 
   
    Each ADO is uniquely identified by the vector $\mathbf{j}$, of length $2(M+1)$, with elements, $j_{nk}$, where $k=0$ corresponds to the primary Brownian oscillator mode for bath $n$ and $k>0$ are Matsubara axes. The frequencies, $\nu_{nk}; k=0,1,2,...,M$, and the coefficients, $c_{nk}$, are,
    \begin{eqnarray}
        \nu_{n0} &=& \Lambda_n,  \label{eq:Over_Mat0} \\
        \nu_{nk} &=& \frac{2\pi k}{\hbar\beta}, \label{eq:Over_Matk} \\
        c_{n0} &=& \eta_n\Lambda_n\left(\cot\left(\frac{\hbar\beta\Lambda_n}{2}\right)-i\right), \label{eq:Over_c0} \\
        c_{nk} &=& \frac{4\eta_n\Lambda_n}{\hbar\beta}\left(\frac{\nu_{nk}}{\nu_{nk}^2-\Lambda_n^2}\right). \label{eq:Over_ck}
    \end{eqnarray}
    The first term in eq.~\eqref{eq:Over_HEOMRate} corresponds to 
    independent propagation of the system and bath, whilst the terms involving $\mathbf{j}^\pm$ are responsible for the interaction between them. The final double commutator term in eq.~\eqref{eq:Over_HEOMRate} is a low-temperature correction which provides additional stability.\cite{Tanimura2006}
    
    The reduced density operator of the system corresponds to $\rho_\mathbf{0}$, with all $j_{nk}=0$, defined as a Boltzmann distribution over the vibrational levels of the ground electronic state prior to a \SI{2}{\ps} equilibration period to achieve correlated initial conditions.\cite{Ishizaki2007,Tanimura2014}     
    
    The hierarchy is terminated by enforcing the Markovian limit as $\xi$, where $\xi \gg \Lambda_n$, such that,
    \begin{equation}
        \frac{2(M+1)\pi}{\hbar\beta}>\xi,
    \end{equation}
    determines the number of Matsubara axes for each bath, $M$, and
    \begin{equation}
        \sum_{n=1}^N\sum_{k=0}^M{{j}_{nk}\nu_{nk}}>\xi,
    \end{equation}
    limits the maximum value of $j_{nk}$ for each axis.\cite{Dijkstra2017, Green2019QuantifyingLineshape} The equation of motion for terminating ADOs is thus assumed to be Markovian, approximated as,
    \begin{equation}
       \dot\rho_\mathbf{j}(t) \simeq -\frac{i}{\hbar}{H}_\text{S}^{\prime\times}\rho_\mathbf{j}(t).
    \end{equation}

    Here we employ $N=2$ baths representing different dissipation processes introduced by the vibrational motion of the solvent environment. The first bath, $n=1$, with the system-bath coupling operator,
    \begin{equation}
        {B}_1=\sum_i\ket{i}\!\bra{i} \otimes Q,
    \end{equation}
    where $i=\{g,e,f\}$, accounts for vibrational relaxation and dephasing for all electronic states, responsible for the decay of vibrational wavepackets, but results in a blueshift of the vibrational frequency of the system which is corrected by adjusting the mode frequency in the system Hamiltonian as $\omega_0= \omega_0^\prime - \eta_1$.\cite{Bennett2018} 
    
    The second bath, $n=2$, with the diagonal coupling operator,
    \begin{equation}
        {B}_2 = (\ket{e}\!\bra{e}+2\ket{f}\!\bra{f})\otimes I_\text{vib},
    \end{equation}
    introduces fluctuations in the electronic transition frequencies responsible for electronic dephasing. The factor of 2 ensures the dephasing rates of S$_1$$\leftarrow$S$_0$ and S$_\text{n}$$\leftarrow$S$_1$ are the same.

\begin{figure}
        \centering
        \includegraphics{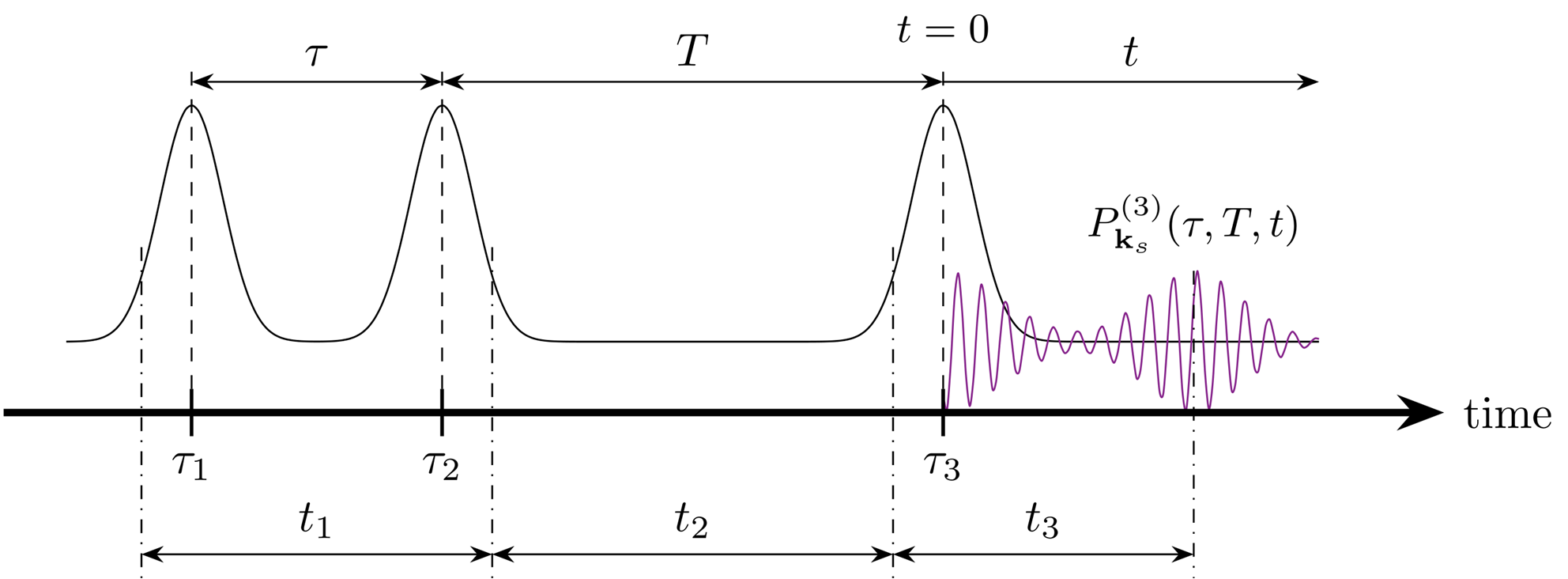}
        \caption{Interaction timeline for 2D electronic spectroscopy. }   
        \label{fig:PulseSequence}
    \end{figure}

    2DES involves three system-field interactions, with the first and second separated by the coherence time, $\tau$, and the second and third by the population time, $T$, producing a third order polarization which dephases over the final time period, $t>0$ (fig. \ref{fig:PulseSequence}).\cite{Biswas2022CoherentSpectroscopies}

    The system-field interaction Hamiltonian in the dipole approximation is split into three portions,\cite{Gelin2005,Gelin2009}
    \begin{equation} 
       {H}_\text{SF}(t) = -\hat{\mu}\mathcal{E}(\mathbf{r},t)
       = -\sum_{m=1}^3{V_m(t)+c.c.}, 
    \end{equation}
    where $c.c.$ is the complex conjugate and,
    \begin{equation}
       V_m(t) =\hat{\mu}\chi_mE^\prime_m(t-\tau_m)\exp(i\textbf{k}_m\cdot\textbf{r}-i\omega_mt),
    \end{equation}
    with frequency $\omega_m=2\pi\nu_m$, wavevector $\mathbf{k}_m$ and electric field strength, $\chi_m$. The field envelope, $E^\prime_m(t-\tau_m)$, centred at $\tau_m$, is assumed to be Gaussian with FWHM $\tau_p$,\cite{Sharp2010,Leng2017a}
    \begin{equation}
        E^\prime_m(t-\tau_m) = \exp\left(\frac{-4\ln2(t-\tau_m)^2}{\tau_p^2}\right).
    \end{equation}
    The first two interactions, $m=1,2$, correspond to the pump pulse, whilst $m=3$ is the interaction with the white light continuum probe pulse. Note that calculation of the spatial phase can be simplified as $\mathbf{k}_m\cdot\mathbf{r}=\omega_m\tau_m$.\cite{Green2018SpectralModel}

    The full influence of the finite pump spectrum is incorporated using the equation-of-motion phase-matching approach (EOM-PMA),\cite{Gelin2022Equation-of-MotionSignals} which involves propagation of seven auxiliaries via eq.~\eqref{eq:Over_HEOMRate} with different Liouvillian operators which correspond to variations in the number and order of pulse interactions:
    \begin{align}
        \mathcal{L}_1\rho_1(t) &= \left[{H}_\text{S}-V_1(t)-V_2^\dagger(t)-V_3^\dagger(t),\rho_1(t) \right], \label{eq:rho1} \\
       \mathcal{L}_2\rho_2(t) &= \left[{H}_\text{S}-V_1(t)-V_2^\dagger(t),\rho_2(t) \right], \label{eq:rho2} \\
        \mathcal{L}_3\rho_3(t) &= \left[{H}_\text{S}-V_1(t)-V_3^\dagger(t),\rho_3(t) \right], \\
        \mathcal{L}_4\rho_4(t) &= \left[{H}_\text{S}-V_1(t),\rho_4(t) \right], \\
        \mathcal{L}_5\rho_5(t) &= \left[{H}_\text{S}-V_2^\dagger(t)-V_3^\dagger(t),\rho_5(t) \right], \\
        \mathcal{L}_6\rho_6(t) &= \left[{H}_\text{S}-V_2^\dagger(t),\rho_6(t) \right], \\
        \mathcal{L}_7\rho_7(t) &= \left[{H}_\text{S}-V_3^\dagger(t),\rho_7(t) \right]. \label{eq:rho7} 
    \end{align}
    Combination of these auxiliaries then isolates the third order polarization produced by a single interaction with each pulse in the phase-matched direction, $\textbf{k}_s$,\cite{Gelin2005}
    \begin{align} \label{eq:Spec:3rd_Pol_EOMPMA}
        P^{(3)}_{\textbf{k}_s}(\tau,T,t)=\text{Tr}\big( & \hat{\mu}(\rho_1(t)-\rho_2(t)-\rho_3(t)+\rho_4(t) \nonumber \\ &  -\rho_5(t)+\rho_6(t)+\rho_7(t)) \big)+c.c.  
    \end{align}
    where the complex conjugate term accounts for the complementary set of Liouville pathways. Note that the number of auxiliaries can be reduced by enforcing the rotating wave approximation,\cite{Cheng2007b,Gelin2009} which is not used here.
   
    The time ordering of the pump interactions yields either the rephasing, $\textbf{k}_\text{R}=-\textbf{k}_1+\textbf{k}_2+\textbf{k}_3$, or nonrephasing, $\textbf{k}_\text{NR}=\textbf{k}_1-\textbf{k}_2+\textbf{k}_3$, polarizations, which correspond to $\tau>0$ or $\tau<0$, respectively, in a noncollinear geometry.\cite{Biswas2022CoherentSpectroscopies} 2D electronic spectra are obtained by Fourier transform with respect to both $\tau$ and $t$,
    \begin{equation}
        S_{\text{R}}(\omega_{\tau},T,\omega_t)=\int^\infty_{0}\text{d}t\int^\infty_{0}\text{d}\tau e^{-i\omega_{\tau}\tau}e^{i\omega_tt}{iP^{(3)}_{\textbf{k}_\text{R}}(\tau,T,t)},
        \label{eq:S_R}
    \end{equation}
    \begin{equation}
        S_{\text{NR}}(\omega_{\tau},T,\omega_t)=\int^\infty_{0}\text{d}t\int^\infty_{0}\text{d}\tau e^{i\omega_{\tau}\tau}e^{i\omega_tt}{iP^{(3)}_{\textbf{k}_\text{NR}}(\tau,T,t)},
        \label{eq:S_NR}
    \end{equation}
    where the reversal in time ordering with respect to the coherence time requires the opposite transform and the sum of rephasing and nonrephasing yields the absorptive spectrum, $S_\text{A}=\text{Re}(S_\text{R}+S_\text{NR})$. 2D spectra then reveal the correlation of the excitation axis, $\omega_\tau$, from the pump interactions with the detection axis, $\omega_t$, from interaction with the probe for each waiting time, $T$.

    PP spectra correspond to the same third order polarization when the two pump interactions occur simultaneously, with $\mathbf{k}_1 = \mathbf{k}_2$ and $\tau=0$. Thus the polarization in the phase-matched direction for PP is similarly calculated using the EOM-PMA as,\cite{Gelin2005} 
   \begin{equation} \label{eq:EOMPMA_PP}
        P^{(3)}_\text{PP}(T,t)=\text{Tr}\big(\hat{\mu}\left( \rho_1(t)-\rho_2(t)-\rho_7(t)\right) \big)+\text{c.c.},
   \end{equation}
    in terms of the same set of auxiliaries defined in eqs.~\eqref{eq:rho1}, \eqref{eq:rho2} and \eqref{eq:rho7}.
    
    PP spectra are then obtained for each $T$ via Fourier transform with respect to the probe axis only, 
    \begin{equation}
        S_{\text{PP}}(T,\omega_t)=-\text{Re}\int^\infty_{0}\text{d}t e^{i\omega_tt}{iP^{(3)}_\text{PP}(T,t)},
        \label{eq:S_PP}
    \end{equation}
    where the factor of $-1$ converts the sign convention from 2DES with ESA $<0$ to PP with ESA $>0$.
    
    In the impulsive limit, in which the field envelope $E^\prime_m(t-\tau_m) = \delta(t-\tau_m)$, spectra are obtained directly from the molecular response function.\cite{Mukamel1995} The steady state absorption spectrum is calculated in the impulsive limit as the Fourier transform of the first order molecular response function,\cite{Chen2009,Tanimura2012}
    \begin{equation} 
        \sigma_A(\omega) = \int^\infty_{0}{\text{d}t e^{i\omega t}\text{Tr}_g\left(\hat\mu G(t,0)[\hat\mu,\rho_\mathbf{j}(-\infty)]\right)},
    \end{equation}
    where the trace is taken over the ground electronic state of $\rho_\mathbf{0}$ only and $G(t,0)$ corresponds to propagation from time $0$ to $t$ using the HEOM, eq.~\eqref{eq:Over_HEOMRate}, after application of the commutator to all equilibrated ADOs. Assuming rapid vibrational relaxation, the steady state emission spectrum for the DHO is then simply the mirror image of the absorption spectrum reflected about $\omega_{eg}=(E_e - E_g)/\hbar$.\cite{Mukamel1995}
    
    Similarly, distinguishing the raising ($\hat\mu_+$) and lowering ($\hat\mu_-$) contributions to the dipole moment operator, the rephasing, $R^{(3)}_\text{R}$, and nonrephasing, $R^{(3)}_\text{NR}$, third order molecular response functions are given by,\cite{Ishizaki2007}
    \begin{widetext}
    \begin{equation}
        R^{(3)}_\text{R}(\tau,T,t)=\text{Tr}\left(\hat\mu G(t+T+\tau,T+\tau)\frac{i}{\hbar}\hat\mu^\times_+ G(T+\tau,\tau)\frac{i}{\hbar}\hat\mu^\times_+ G(\tau,0)\frac{i}{\hbar}\hat\mu^\times_-\rho_\mathbf{j}(-\infty)\right),
    \end{equation}
    \begin{equation}
        R^{(3)}_\text{NR}(\tau,T,t)=\text{Tr}\left(\hat\mu G(t+T+\tau,T+\tau)\frac{i}{\hbar}\hat\mu^\times_+ G(T+\tau,\tau)\frac{i}{\hbar}\hat\mu^\times_- G(\tau,0)\frac{i}{\hbar}\hat\mu^\times_+\rho_\mathbf{j}(-\infty)\right),
    \end{equation}
    \end{widetext}
    and transformed into the rephasing and nonrephasing 2D spectra via eq.~\eqref{eq:S_R} and \eqref{eq:S_NR}, respectively, where $P^{(3)}=R^{(3)}$.

    PP spectra in the impulsive limit are similarly obtained from the third order molecular response function,\cite{Zhang2021ProbingSpectroscopy}
    \begin{equation} \label{eq: R_PP}
        R^{(3)}_\text{PP}(T,t)=\text{Tr}\left( \hat\mu G(t+T,T) \frac{i}{\hbar}\hat\mu_+^\times G(T,0) \hat\rho^{(2)}(0) \right),
    \end{equation}
    where $\hat\rho^{(2)}(0) = \hat\rho^{(2)}_\text{R}(0) + \hat\rho^{(2)}_\text{NR}(0)$, which directly produces the absorptive spectrum via summation of rephasing and nonrephasing pathways for $\tau=0$,
    \begin{equation}\label{rho2(0)_reph}
        \hat\rho^{(2)}_\text{R}(0) = -\hat\mu_{+}^\times\hat\mu_{-}^\times\hat\rho_\mathbf{j}(-\infty)/\hbar^2,
    \end{equation}
    \begin{equation}\label{rho2(0)_nonreph}
        \hat\rho^{(2)}_\text{NR}(0) = -\hat\mu_{-}^\times\hat\mu_{+}^\times\hat\rho_\mathbf{j}(-\infty)/\hbar^2,
    \end{equation}
    respectively, and is transformed via eq.~\eqref{eq:S_PP}. Note that the sum of rephasing and nonrephasing pathways, rather than commutation with the total dipole moment operator, is required to avoid including two quantum (2Q) pathways involving an $\ket{f}\!\bra{g}$ coherence, which were not an issue for the two electronic state system for which this approach was previously described in ref.~\citenum{Zhang2021ProbingSpectroscopy}.

     We base our model on the most intense Raman active mode of the well-known dye molecule cresyl violet for which $\omega_{eg}/{2\pi c}=\SI{16260}{\per\cm}$, $\omega_{fe}/{2\pi c}=\SI{19940}{\per\cm}$ and ${\omega_0^\prime}/{2\pi c}$ = \SI{585}{\per\cm}.\cite{Bressan2024Two-DimensionalDisplacements} Impulsive stimulated Raman spectroscopy has determined the S$_1$$\leftarrow$S$_0$ displacement for this mode to be $\Delta_{eg}=0.63$.\cite{Batignani2021Excited-StateProfiles} The diabatic system Hamiltonian is constructed with 10 vibrational levels per electronic state prior to diagonalization into the adiabatic basis and truncation to 5 vibrational levels to reduce computation time. The dissipation rate of both baths is set at $\Lambda_1 = \Lambda_2$ = \SI{160}{\per\cm}, which corresponds to a correlation lifetime of $\Lambda_n^{-1}=\tau_c \approx \SI{208}{\fs}$. Weak coupling to the vibrational bath, ${\eta_1}/{2\pi c}=\SI{40}{\per\cm}$, results in the slow, homogeneous, decay of vibrational wavepackets such that beatings survive beyond $T = \SI{1}{\ps}$. Stronger coupling to the electronic bath, ${\eta_2}/{2\pi c}=\SI{300}{\per\cm}$, introduces significant inhomogeneous broadening that overwhelms the vibronic progression and induces a large Stokes shift approaching $2\eta_2$ with rapid spectral diffusion.\cite{Green2019QuantifyingLineshape} For these parameters, setting the Markovian limit to $\xi = 10\Lambda_1$ produces a hierarchy with 355 ADOs. 

\section{Results and Discussion}

\begin{figure*}
    \centering
    \includegraphics{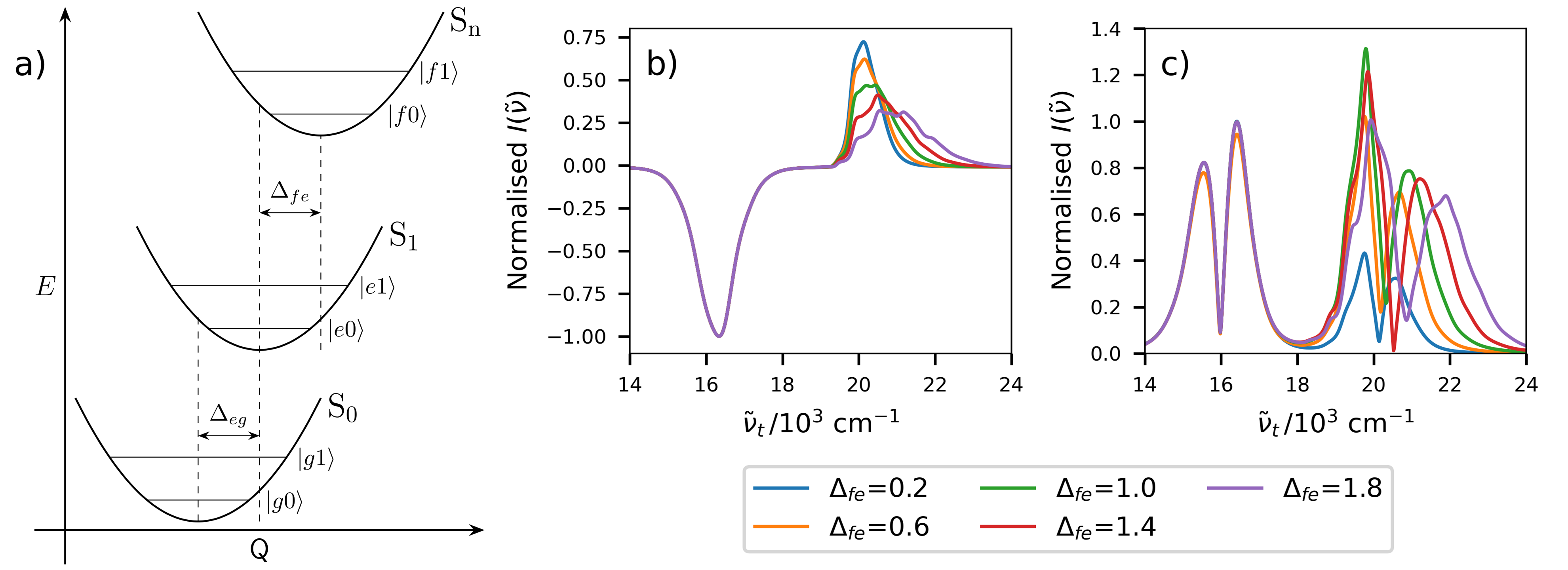}
    \caption{(a) PESs for three-state DHO model. (b) Impulsive PP spectra at $T=\SI{2}{\ps}$ and (c) FCS at $\tilde\nu_T=\SI{585}{\wn}$ for increasing $\Delta_{fe}$, normalised to $\Delta_{fe}=0.2$.}
    \label{fig:PES-PP-FCS}
\end{figure*}

  A schematic diagram for the PESs of the three-state DHO model is shown in fig.~\ref{fig:PES-PP-FCS}(a). Fig.~\ref{fig:PES-PP-FCS}(b) shows PP spectra calculated in the impulsive limit for a series of S$_\text{n}$$\leftarrow$S$_1$ displacements in the range $\Delta_{fe}$ = 0.2--1.8, normalised to the intensity of the negative peak, which corresponds to ground state bleach (GSB) and  stimulated emission (SE) from S$_1$ and is unchanged due to the fixed $\Delta_{eg}$ = 0.63. Note that whilst our general model allows for negative displacements, this range of $\Delta_{fe}$ assumes the equilibrium position increases for higher energy excited states and includes very large values of $\Delta_{fe}>1$ to assess the trends over an extended range. Increasing $\Delta_{fe}$ results in broadening of the ESA band associated with an expanding vibrational progression as the excited state (S$_1$) wavepacket is projected vertically onto higher vibrational levels of the S$_\text{n}$ potential by the probe interaction. 

  Fig.~ \ref{fig:PES-PP-FCS}(c) shows the corresponding FCS for increasing $\Delta_{fe}$, calculated as the Fourier transform of the PP residuals after global fit along the population time, $T$.\cite{Arpin2021SignaturesSpectra,Barclay2022CharacterizingSpectra} In the impulsive limit, FCS relate to beatings of the excited state (S$_1$) wavepacket only (see later discussion of fig.~\ref{fig:pump_FCS}),\cite{Banin1994ImpulsivePulses,McClure2014CoherentState} where superposition of the vibrational wavefunctions with alternating even and odd parity results in a node in both the lower frequency SE peak at the S$_0$$\leftarrow$S$_1$ emission frequency and at the  S$_\text{n}$$\leftarrow$S$_1$ absorption frequency in the higher frequency ESA band.\cite{Turner2020BasisSpectra, Fitzpatrick2020SpectralExcitation} The asymmetry of the spectra either side of the node for both the SE and ESA peaks reflects the excited state displacements of the harmonic PESs.\cite{Arpin2021SignaturesSpectra} On increasing $\Delta_{fe}$ the ESA node blue shifts, following the blue shift in the S$_\text{n}$$\leftarrow$S$_1$ absorption maximum in fig.~\ref{fig:PES-PP-FCS}(b). When $\Delta_{fe}<\Delta_{eg}$ the amplitude of the ESA band is less than that of the SE band. When $\Delta_{fe}\approx\Delta_{eg}$ the ESA and SE bands have roughly equal intensity. When $\Delta_{fe}>\Delta_{eg}$,  the amplitude of the ESA band initially exceeds that of the SE, until the expanding vibrational progression broadens the ESA band sufficiently that the maximum amplitude begins to decrease; seen at $\Delta_{fe}>1$ in fig.~\ref{fig:PES-PP-FCS}(c). These observations would naturally be scaled if $\mu_{eg}\ne\mu_{fe}$, but are remarkable considering both the SE and ESA bands correspond to probing of the same S$_1$ wavepacket. The relative excited state displacement $\Delta_{fe}/\Delta_{eg}$ therefore has important consequences for the appearance of FCS spectra and can be used as a key measure for the interpretation of ESA coherences.

\begin{figure*}
    \centering
    \includegraphics{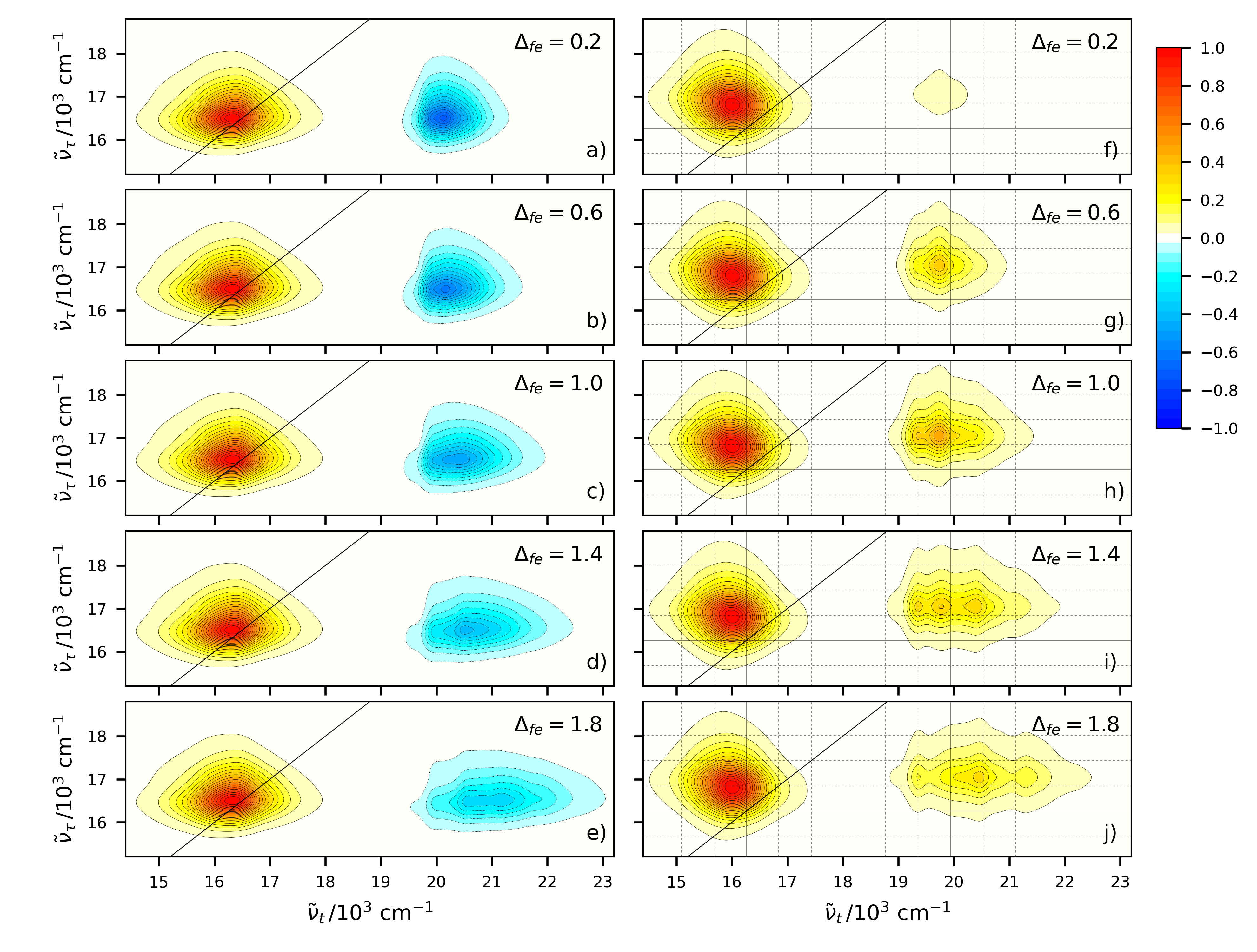}
    \caption{(a--e) Absorptive 2D spectra for $T=\SI{500}{\fs}$ and (f--j) rephasing $+\SI{585}{\wn}$ beat maps for increasing $\Delta_{fe}$, individually normalised.}
    \label{fig:2DES-RephPstv}
\end{figure*}

    Absorptive 2DES spectra, in the impulsive limit, with the same $\Delta_{fe}$ values as in fig. \ref{fig:PES-PP-FCS}(b) are presented in fig. \ref{fig:2DES-RephPstv}(a--e). Similarly to the PP spectra, these show a decrease in the intensity of the negative ESA band as it broadens along the detection (probe, $\tilde{\nu}_t$) axis due to the expanding vibrational progression. 2D beat maps are calculated in an analogous way to FCS; as transforms of the residuals following global fit along $T$. However in the case of 2DES, separate transforms of the rephasing and nonrephasing components distinguish coherence pathways oscillating with positive and negative phase into different beat maps at $\tilde\nu_T = \pm\SI{585}{\wn}$.\cite{Seibt2013,Green2018SpectralModel} These beat maps do not feature the nodes observed in FCS which are the result of phase cancellation because absorptive PP spectra are the result of the summation of rephasing and nonrephasing coherence pathways.\cite{Dean2015BroadbandBlue} The corresponding rephasing $\tilde\nu_T = +\SI{585}{\wn}$ beat maps in fig. \ref{fig:2DES-RephPstv}(f--j) show the same expanding progression in the ESA band along the detection axis from fig. \ref{fig:PES-PP-FCS}(c). The maximum amplitude of the ESA band also increases from fig. \ref{fig:2DES-RephPstv}(f)--(g)--(h) before decreasing from (h)--(i)--(j), as in fig. \ref{fig:PES-PP-FCS}(c) with increasing $\Delta_{fe}$, with the initial increase as $\Delta_{fe}$ surpasses $\Delta_{eg}$ and the subsequent decrease due to the amplitude spread over the broader progression. However, unlike fig. \ref{fig:PES-PP-FCS}(c), the maximum amplitude of the ESA band never exceeds that of the lower detection frequency band as rephasing positive is one of four beat maps for the \SI{585}{\wn} mode, the others being rephasing negative, and non-rephasing positive and negative. Consequently, only a fraction of the total pathways contribute to spectra in fig \ref{fig:2DES-RephPstv}(f--j) compared to the FCS. Nevertheless, the significant change in intensity of the ESA coherences in the beat maps on increasing $\Delta_{fe}$ again highlights the impact of the relative excited state displacement $\Delta_{fe}/\Delta_{eg}$ in resolving ESA coherences, even in the impulsive limit.

\begin{figure}
    \centering
    \includegraphics{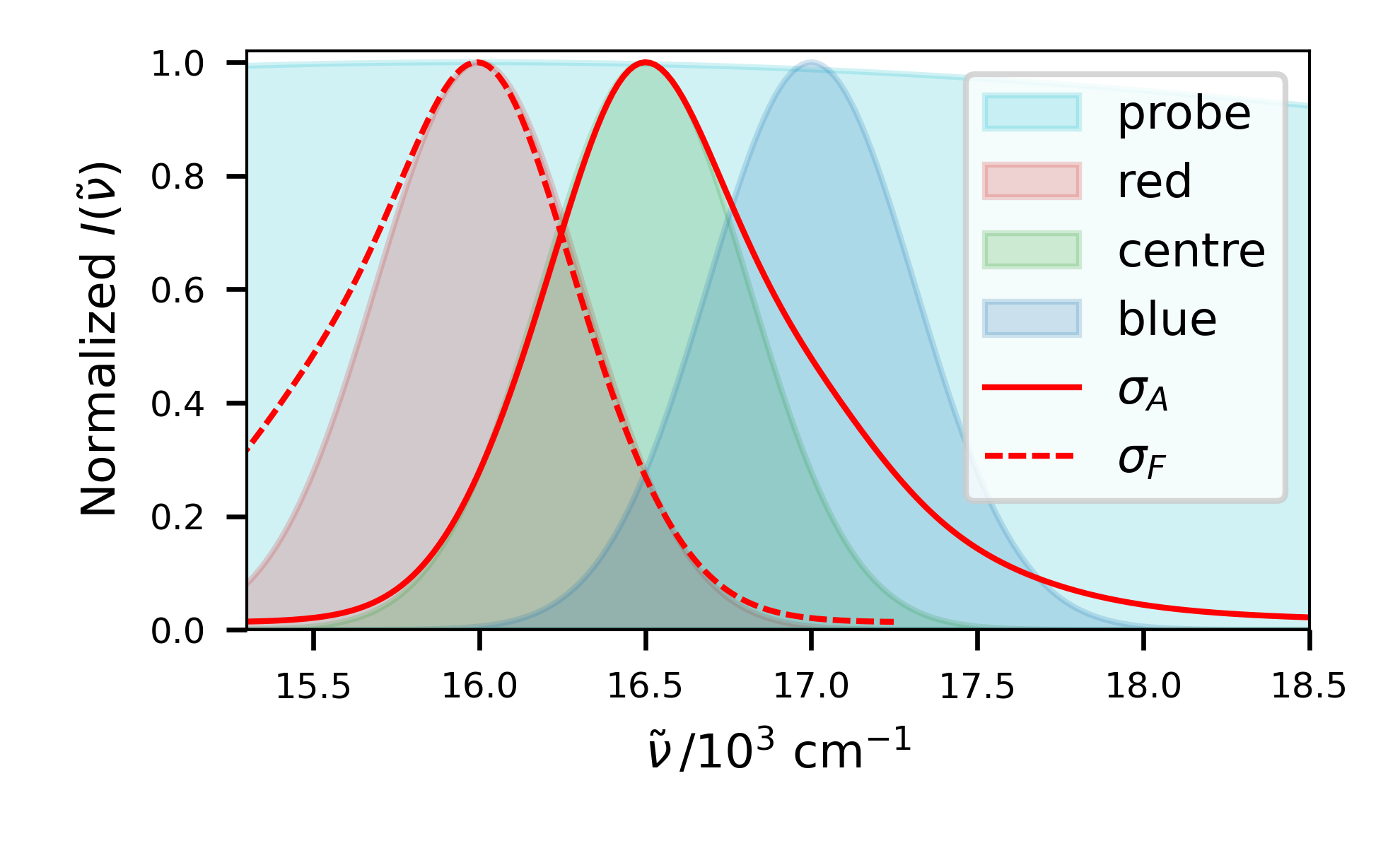}
    \caption{Pump (red, centre and blue shifted) and probe spectra relative to the steady state absorption, $\sigma_A$, and fluorescence, $\sigma_F$, spectra.}
    \label{fig:pump_spectra}
\end{figure}

    A finite pump spectrum, as is typical in experiments, will obscure coherences by excluding pathways that involve transition frequencies beyond the limits of the laser spectrum, in a phenomenon known as spectral filtering.\cite{Green2018SpectralModel} To demonstrate this, we henceforth fix $\Delta_{fe} = 0.6$ and show the impact of a series of different finite pump spectra on the FCS and 2D beat maps. We consider laser spectra that are red, centre and blue shifted, compared to the S$_1$$\leftarrow$S$_0$ transition of the molecule.   
    The pump pulse is modelled using a Gaussian with FWHM of $\tau_p =  \SI{40}{\fs} $, equivalent to $\tilde\nu_p$ = \SI{736}{\wn}, with $\omega_{1,2} = $\SI{16000}{\wn}, \SI{16500}{\wn} or \SI{17000}{\wn} corresponding to a red shifted, centre or blue shifted spectrum with respect to the steady state absorption spectrum, respectively (fig.~\ref{fig:pump_spectra}) and field strength  $\chi_{1,2} = \SI{1E7}{\V\per\m}$. The weaker white light continuum probe pulse is represented by a Gaussian approaching the impulsive limit with $\tau_p = \SI{2}{\fs}$, centred at $\omega_{3} = \SI{16000}{\per\cm}$ with field strength of $\chi_{3} =  \SI{1E5}{\V\per\m}$. The polarization vectors of the pulses are assumed parallel to the electronic transition dipole moments of the system.

\begin{figure*}
    \centering
    \includegraphics{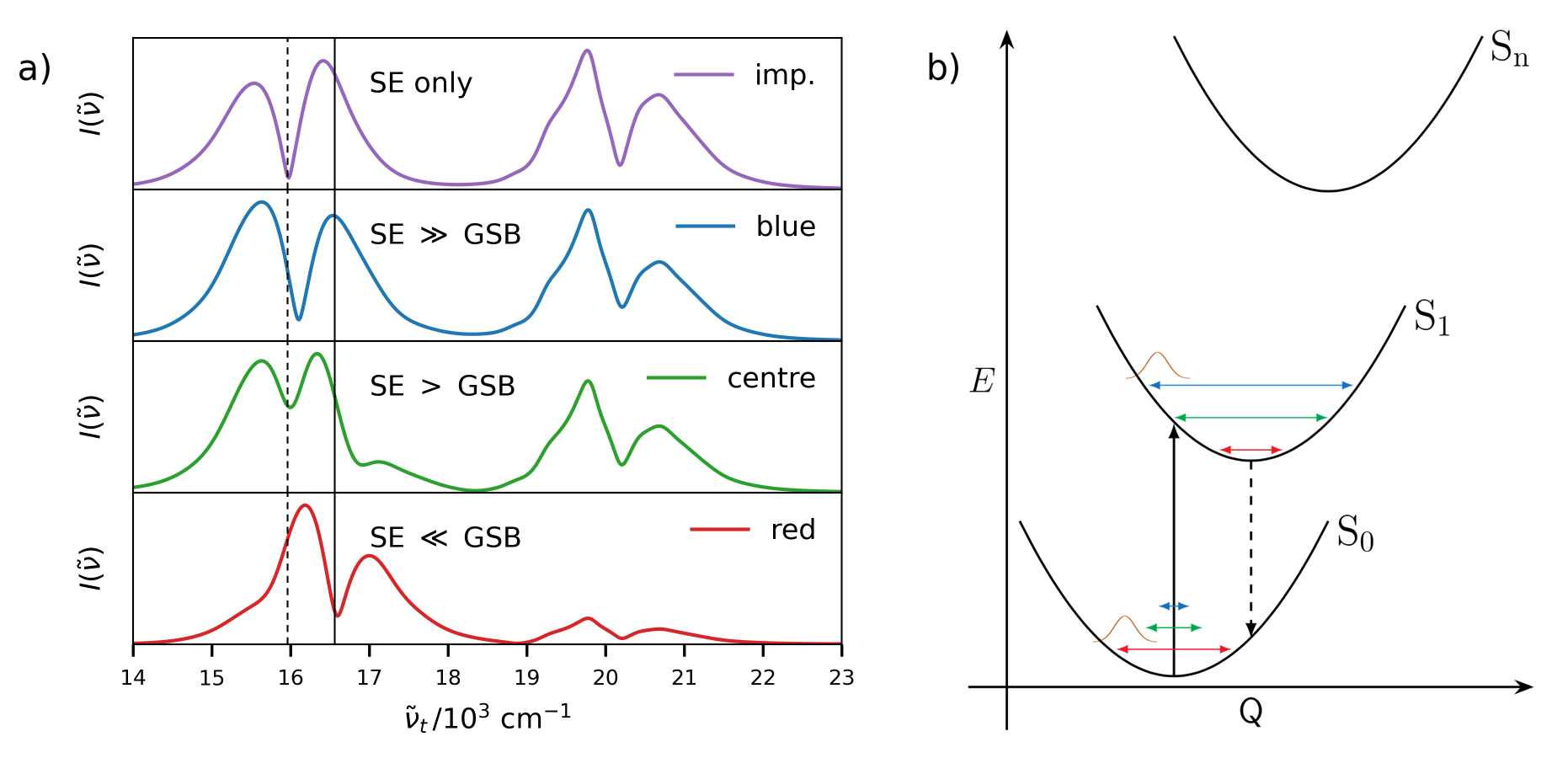}
    \caption{(a) FCS in the impulsive limit compared with FCS using the blue, centre and red shifted pump spectra shown in fig.~\ref{fig:pump_spectra} for $\Delta_{fe}=0.6$. The solid (dashed) black line corresponds to the absorption (emission) maximum identified by the arrow on the PES in (b).} 
    \label{fig:pump_FCS}
\end{figure*}

    Fig.~\ref{fig:pump_FCS}(a) shows FCS calculated in the impulsive limit (purple) vs.\ the red, centre and blue shifted finite pump spectra. The solid (dashed) line corresponds to the steady state absorption (fluorescence) maximum identified by the solid (dashed) arrow in the PES in fig.~\ref{fig:pump_FCS}(b). The shifting pump spectrum causes the intensity and lineshape of the lower detection frequency ($\tilde\nu_t$) GSB+SE peak to vary significantly, whilst the ESA lineshape and node remain unchanged, determined by $\Delta_{fe}$ as discussed above. The changing lineshape and node position in the GSB+SE is explained by evolution of the ground (S$_0$) vs.\ excited (S$_1$) state wavepackets within the time frame of the interactions.\cite{Kumar2001InvestigationsPulses,Zhu2020ElectronicSpectroscopy,Fitzpatrick2020SpectralExcitation} In the impulsive limit, the two instantaneous pump interactions simultaneously excite and return a ground state population to its equilibrium position. Consequently there is no time to produce an oscillatory ground state wavepacket that contributes GSB coherence pathways to the FCS.\cite{Pollard1990Quantum-mechanicalBlue,Banin1994ImpulsivePulses,Jonas1995FemtosecondDuration} The impulsive limit FCS thus shows SE only from the excited state wavepacket. This results in a node which corresponds to the minimum of the S$_1$ potential (i.e. the maximum of the steady state emission spectrum) identified by the dashed line/arrow in fig.~\ref{fig:pump_FCS}. However, for non-impulsive pump spectra the finite duration of the two pump interactions allows the excited wavepacket to gain momentum in the excited state potential prior to projection back down to the ground state.\cite{Pollard1992TheoryBacteriorhodopsin,Bardeen1998FemtosecondBacteriorhodopsin} Upon returning to the ground state, the wavepacket is displaced from equilibrium and oscillates on the S$_0$ potential, contributing GSB coherence pathways to the FCS.\cite{Banin1994ImpulsivePulses, Rafiq2016} The turning points of the ground and excited state wavepackets are indicated by the double-headed arrows on the PES in fig.~\ref{fig:pump_FCS}(b), with colours corresponding to the associated pump spectrum. Excited state wavepackets traverse the entire width of the S$_1$ harmonic potential, whilst ground state wavepackets oscillate about the ground state minimum, limited by the momentum gained in returning from the S$_1$ surface during the finite pump interaction.\cite{Pollard1990Quantum-mechanicalBlue,Pollard1990WaveExperiments,Bardeen1998FemtosecondBacteriorhodopsin} The lengths of these arrows qualitatively indicate the amplitude of the wavepacket oscillations and the relative intensity of the GSB vs.\ SE  contributions to the FCS. 
    
    In the case of the centre pump spectrum (green), the GSB amplitude is increased, relative to the impulsive limit case (purple), by the finite pump duration. Weakening of the node at the fluorescence maximum is accompanied by emergence of a third peak at ca.\ $\tilde\nu_t=\SI{17000}{\per\cm}$ in the FCS in fig.~\ref{fig:pump_FCS}(a) which shows contributions from both GSB and SE, where SE is greater. On blue shifting the pump spectrum, population is projected further up the S$_1$ potential, increasing the amplitude of the excited state wavepacket whilst decreasing that of the ground state, such that the blue FCS is almost entirely SE, with the small GSB contribution resulting in a blue shift of the node towards the steady state absorption maximum. Conversely, on red shifting the pump spectrum, population is projected further down the S$_1$ potential, decreasing the amplitude of the excited state wavepacket whilst increasing that of the ground state such that GSB dominates the red FCS, with the node at the steady state absorption maximum (solid line/arrow). Furthermore, when SE dominates (purple/blue), the intensities of the ESA and SE bands are similar due to the choice of $\Delta_{eg}\approx\Delta_{fe}$, as discussed above. Consequently, domination of the GSB for the red pump spectrum significantly reduces the relative intensity of the ESA in the red FCS.

\begin{figure*}
    \centering
    \includegraphics{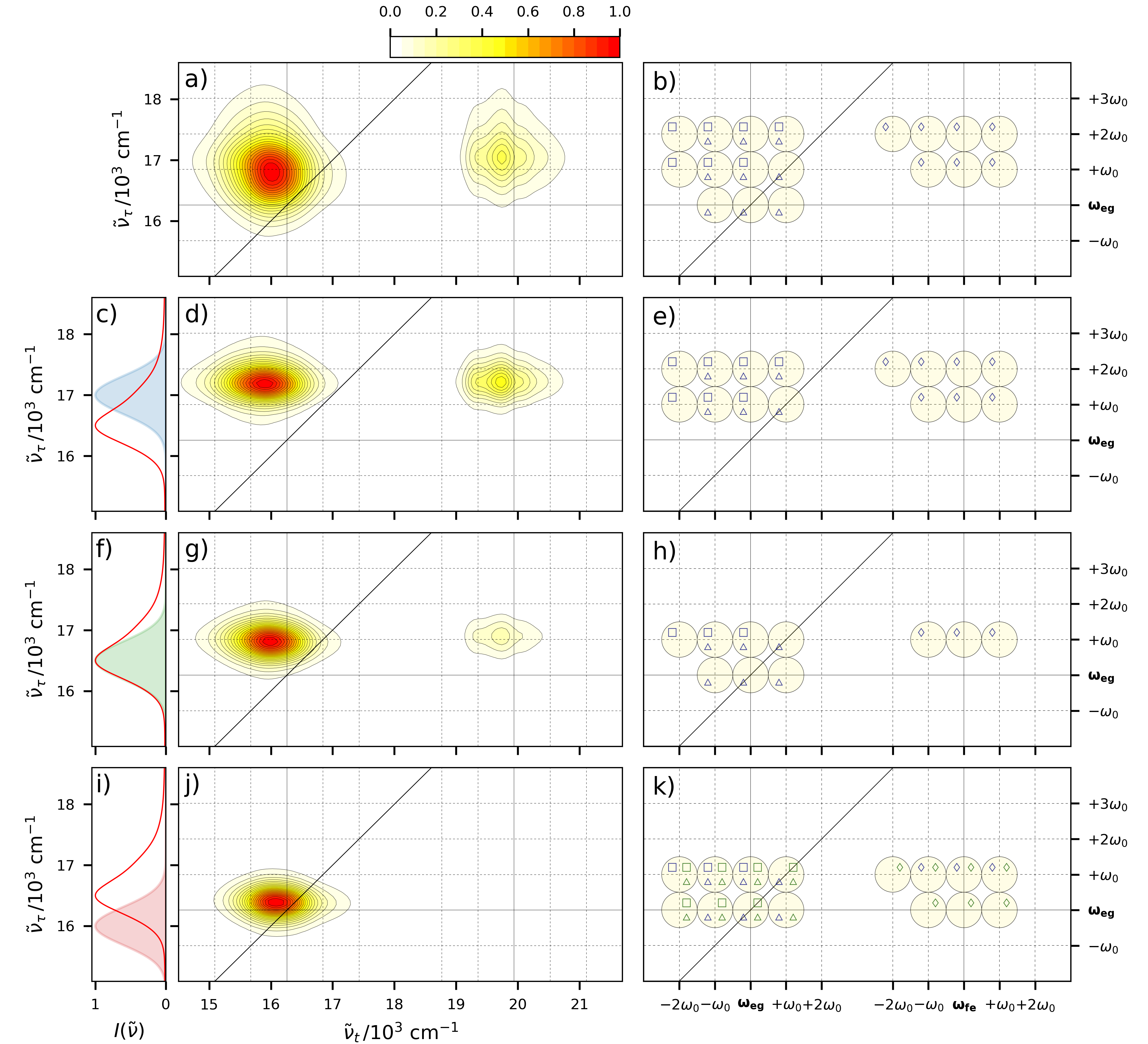}
    \caption{(a, d, g, j) Rephasing $+\SI{585}{\wn}$ beat maps and (b, e, h, k) corresponding location key diagrams for the impulsive limit, blue shifted, centre and red shifted pump spectra, respectively. (c) Blue shifted, (f) centre and (i) red shifted pump spectra (shaded) overlaid with the steady state absorption spectrum (red line).}
    \label{fig:pump_RephPstv}
\end{figure*}

    Similar effects are also understood by considering the associated 2D beat maps, which spread the FCS over the excitation axis and separate the rephasing/nonrephasing positive/negative frequency pathways. 2D spectra were calculated in steps of \SI{20}{\fs} up to $T = \SI{1}{\ps}$. In an approach analogous to those used in experiments, non-oscillatory population pathways were removed via global fit along the population time, $T$, which required two exponential decay components, one fast (ca. \SI{50}{\fs}) and one slow ($>\SI{1}{\ns}$), as well as trimming of the first $T=\SI{100}{\fs}$ to remove distortion at early times due to the coherent artefact. Separate Fourier transform of the rephasing and nonrephasing residuals then produces vibrational beat maps which distinguish positively, $\propto \exp{(+i\omega_0 T)}$, and negatively, $\propto \exp{(-i\omega_0 T)}$, oscillating coherence pathways.\cite{DeA.Camargo2017,Green2018SpectralModel} Analysis of the Liouville pathways for the three state DHO system enables creation of a location key diagram for the assignment of each beat map, where the coherence pathways are labelled using colour-coded symbols. GSB pathways are identified by triangles, SE pathways by squares and ESA pathways by diamonds. Positive (negative) coherence pathways starting from $\ket{g0}\!\bra{g0}$ are coloured blue (red), whilst pathways starting from $\ket{g1}\!\bra{g1}$ are coloured green (purple); as per table S1. Double-sided Feynman diagrams for all the labelled pathways are presented in the SI.

    Fig.~\ref{fig:pump_RephPstv}(a) shows the $+\SI{585}{\per\cm}$ rephasing beat map in the impulsive limit. Unlike in PP, separation of the two pump interactions and resolution of the excitation axis in 2DES allows GSB coherence pathways such that there is a lower detection frequency GSB+SE peak and higher detection frequency ESA peak.  The location of these peaks is explained by the corresponding location key diagram in fig.~\ref{fig:pump_RephPstv}(b). This identifies the location on the 2D map of all the contributing Liouville pathways for pump frequencies in the range $\omega_{eg}-\omega_0$ to $\omega_{eg}+2\omega_0$ and probe frequencies in the range $\omega_{eg}\pm2\omega_0$. Note that all pathways do not contribute equally as their intensities are weighted by the Franck-Condon factors involved, and as spectral broadening applies to each pathway, the peak maxima generally appear between the pathway locations. 

    The positive rephasing beat map for the blue shifted pump pulse, fig.~\ref{fig:pump_RephPstv}(d) with the aligned pump spectrum in fig.~\ref{fig:pump_RephPstv}(c), has a GSB+SE peak which is significantly narrower in the excitation axis ($\tilde\nu_{\tau}$) caused by exclusion of GSB pathways excited at the S$_1$$\leftarrow$S$_0$ transition frequency $\omega_{eg}$. The blue shifted pump now concentrates intensity for pathways with excitation frequencies of $\omega_{eg}+\omega_0$ and $\omega_{eg}+2\omega_0$ such that the GSB+SE peak maximum appears between these frequencies in line with the ESA peak maximum. Conversely, the centre pump spectrum in fig.~\ref{fig:pump_RephPstv}(f) excludes pathways at the higher excitation frequency of $\omega_{eg}+2\omega_0$, producing a beat map with a GSB+SE peak in fig.~\ref{fig:pump_RephPstv}(g) of a similar size to fig.~\ref{fig:pump_RephPstv}(d) but now red shifted along the excitation axis. The ESA peak in fig.~\ref{fig:pump_RephPstv}(g) has also reduced in size due to this spectral filtering. The alignment of the GSB+SE peak in fig.~\ref{fig:pump_RephPstv}(g)  with the squares along $\omega_{eg}+\omega_0$ in fig.~\ref{fig:pump_RephPstv}(h) clearly indicates that SE dominates, as is the case for the equivalent FCS in fig.~\ref{fig:pump_FCS}. Finally, the red shifted pump spectrum, fig.~\ref{fig:pump_RephPstv}(i), results in further narrowing and red shifting of the GSB+SE peak, fig.~\ref{fig:pump_RephPstv}(j), where the position of peak maximum and absence of ESA peak shows the dominance of the GSB pathways from the ground state wavepacket, as seen in the FCS of fig.~\ref{fig:pump_FCS}.

\begin{figure*}
    \centering
    \includegraphics{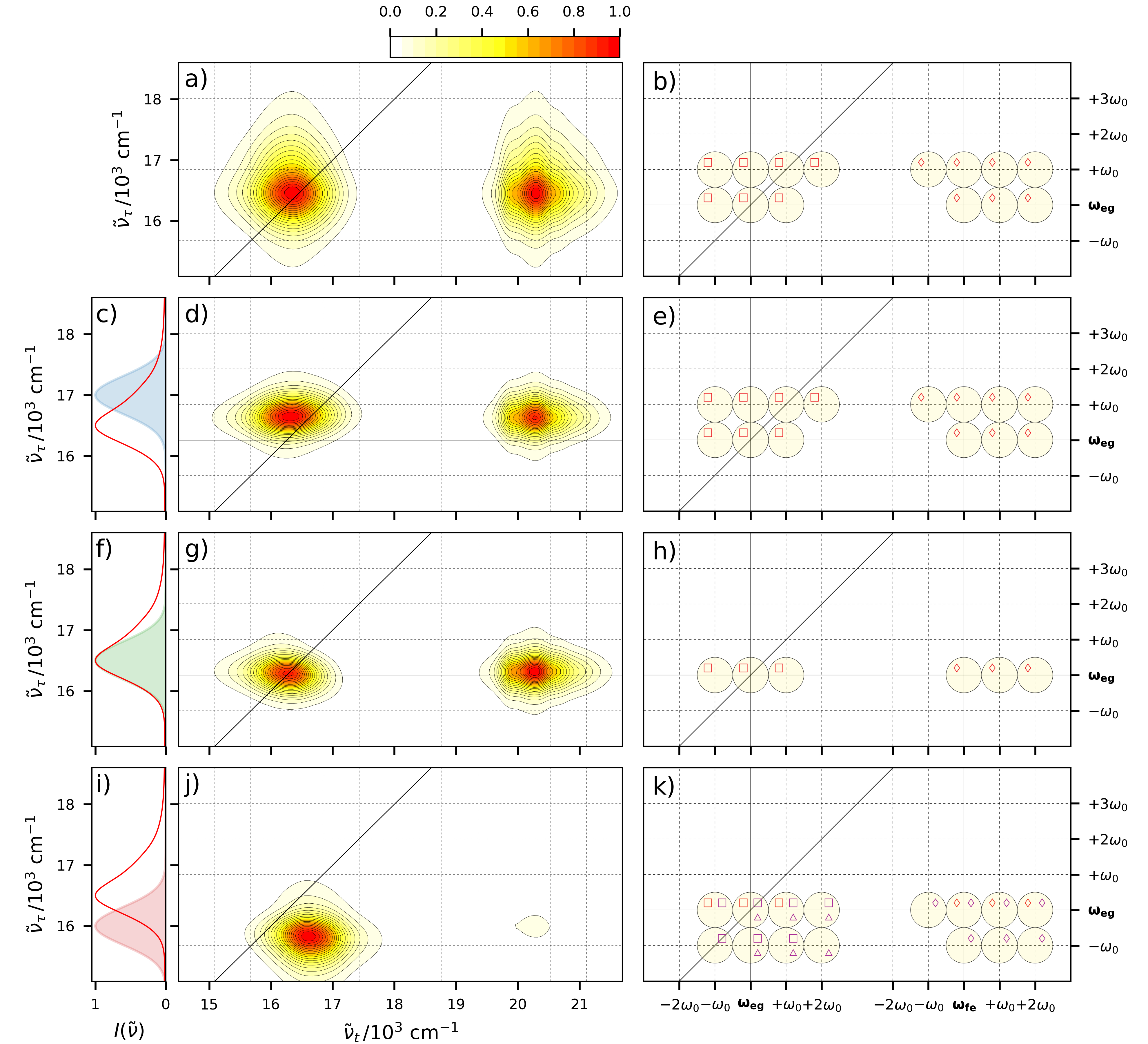}
    \caption{(a, d, g, j) Rephasing $-\SI{585}{\wn}$ beat maps and (b, e, h, k) corresponding location key diagrams for the impulsive limit, blue shifted, centre and red shifted pump spectra, respectively. (c) Blue shifted, (f) centre and (i) red shifted pump spectra (shaded) overlaid with the steady state absorption spectrum (red line).}
    \label{fig:pump_RephNgtv}
\end{figure*}
    
    The negative rephasing beat map in the impulsive limit,  fig.~\ref{fig:pump_RephNgtv}(a), also has two broad peaks, with a higher detection frequency peak corresponding to ESA pathways but a lower detection frequency peak corresponding to SE pathways only, as labelled in fig.~\ref{fig:pump_RephNgtv}(b). As for the positive beat maps, the broadening of these peaks with respect to the excitation axis ($\tilde{\nu}_\tau$) is significantly reduced for the finite, blue shifted pump spectrum in fig.~\ref{fig:pump_RephNgtv}(d), but with negligible change in peak maximum, whilst the centre pump spectrum results in a red shift of SE and ESA peaks in the excitation axis such that both are centred at the S$_1$$\leftarrow$S$_0$ transition frequency $\omega_{eg}$ in fig.~\ref{fig:pump_RephNgtv}(g). The SE peak maximum for the negative rephasing beat map with the red shifted pump spectrum in fig.~\ref{fig:pump_RephNgtv}(j) appears below the S$_1$$\leftarrow$S$_0$ transition frequency in the excitation axis, resulting from Liouville pathways starting from the hot vibrational state $\ket{g1}\!\bra{g1}$, indicated by the purple pathways in fig.~\ref{fig:pump_RephNgtv}(k). The weak but visible ESA peak in fig.~\ref{fig:pump_RephNgtv}(j) also appears below the S$_1$$\leftarrow$S$_0$ transition frequency in the excitation axis showing the domination of these hot band pathways for the red shifted pump spectrum. However, the equivalent hot band pathways for the rephasing positive beat map, labelled in green in fig.~\ref{fig:pump_RephPstv}(k), are a vibrational quantum higher in the excitation axis and contribute much less when excited by the red shifted pump spectrum in fig.~\ref{fig:pump_RephPstv}(i). Note that these hot band pathways also contribute to the impulsive spectra but are obscured by the intensity of pathways starting from the ground vibrational level which has a much larger population at \SI{298}{\K}. The prominence of hot band pathways is therefore achieved by the filtering of the red shifted pump spectrum.

    Similar trends are observed for the nonrephasing 2D beat maps included in the SI. The nonrephasing beat maps again demonstrate spectral filtering as a narrowing and shifting of the peaks in the excitation axis compared with the impulsive spectrum, as well as the appearance of hot band pathways starting from $\ket{g1}\!\bra{g1}$ in the positive nonrephasing beat map for the red shifted pump spectrum in fig.~S1(j). Therefore the selectivity of the excited vs.\ ground state wavepacket observed in the FCS of fig.~\ref{fig:pump_FCS} on shifting the location of the pump spectrum relative to the steady state absorption is also observed in the 2D beat maps, where the transition from dominance of SE for a blue shifted pump to GSB for a red shifted pump is emphasised by the red shift in the peak along the excitation axis of the beat map. 
    
\section{Conclusions}    
  
    The impacts of the displacement of the electronic excited states relative to each other and the spectral window of the pump pulse in HB2DES have been explored using a comprehensive model based upon cresyl violet, which fully accounts for the finite pump spectrum and the solvent environment. 
    Impulsive limit PP has shown that increasing $\Delta_{fe}$ for a fixed $\Delta_{eg}$ results in an increase in ESA amplitude in FCS which exceeds that of GSB+SE when $\Delta_{fe}>\Delta_{eg}$, but is limited by the broadening vibronic progression. A similar progression is observed for 2D beat maps in the impulsive limit. As shown experimentally in ref.~\citenum{Bressan2024Two-DimensionalDisplacements}, this highlights that resolution of ESA coherences in 2D beat maps is determined by the relative excited state displacement $\Delta_{fe}/\Delta_{eg}$ rather than the value of $\Delta_{fe}$ alone. 
    FCS calculated with a finite width pump spectrum then demonstrated the expected trend that when the pump is blue shifted with respect to the steady state absorption spectrum, the FCS is dominated by SE, reflecting the larger amplitude oscillations of the S$_1$ wavepacket, whereas when the pump is red shifted, the FCS is dominated by GSB, corresponding to larger amplitude oscillations of the S$_0$ wavepacket. This transfer of the dominant contribution from SE to GSB upon red shifting the pump spectrum across the absorption band is also observed in 2D beat maps, where it is additionally resolved in the red shift of peaks along the excitation axis. Moreover, the increase in GSB intensity on red shifting the pump spectrum obscures ESA coherences and amplifies the contribution of pathways initially in vibrational hot states, even for the \SI{585}{\wn} mode of cresyl violet at room temperature. Therefore even small shifts in the pump spectrum of a few hundred wavenumbers have a significant impact on FCS and 2D beat maps, presenting an opportunity to tune the intensity of ESA coherences via control of the pump spectrum. 
    The relative displacements of multiple excited states revealed by ESA coherences in HB2DES are potentially very valuable in photophysics and photochemistry and will be increasingly accessible with the development of new ultrabroadband methods, providing modern tools for the investigation of coherences in chemical systems.
     
\section*{SUPPORTING INFORMATION}
    Supporting information includes the nonrephasing 2D beat maps and all double-sided Feynman diagrams.
  
\begin{acknowledgments}

    The research presented in this paper was carried out on the High Performance Computing Cluster supported by the Research and Specialist Computing Support service at the University of East Anglia. We acknowledge support from the Engineering and Physical Sciences Research Council under Awards No. EP/V00817X/1. For the purpose of open access, the authors have applied a Creative Commons attribution (CC BY) licence to any Author Accepted Manuscript version arising.
    
\end{acknowledgments}

\section*{Conflicts of interest}
The authors have no conflicts of interest to disclose. 

\section*{Author Contributions}
\textbf{Dale Green}: Conceptualization (lead), methodology, formal analysis, data curation, software, visualization, writing - original draft.
\textbf{Giovanni Bressan}: Conceptualization (supporting), writing - review and editing (supporting).
\textbf{Ismael A. Heisler}: Writing - review and editing (supporting).
\textbf{Stephen R. Meech}: Funding acquisition (PI), writing - review and editing (supporting).
\textbf{Garth A. Jones}: Funding acquisition (Co-I), writing - review and editing (lead).

\section*{Data Availability}
The data that support the findings of this study are openly available in Zenodo at http://doi.org/10.5281/zenodo.10954929.

\section*{References}
\bibliography{references}

\end{document}


\title{Vibrational coherences in half-broadband 2D electronic spectroscopy: spectral filtering to identify excited state displacements -- Supplementary Information}

\author{Dale Green}
\affiliation{Physics, Faculty of Science, University of East Anglia, Norwich Research Park, Norwich, NR4 7TJ, UK}

\author{Giovanni Bressan}
\affiliation{School of Chemistry, University of East Anglia, Norwich Research Park, Norwich, NR4 7TJ, UK}

\author{Ismael A. Heisler}
\affiliation{Instituto de Física, Universidade Federal do Rio Grande do Sul, 91509-900 Porto Alegre, RS, Brazil.}

\author{Stephen R. Meech}
\affiliation{School of Chemistry, University of East Anglia, Norwich Research Park, Norwich, NR4 7TJ, UK}

\author{Garth A. Jones}
\affiliation{School of Chemistry, University of East Anglia, Norwich Research Park, Norwich, NR4 7TJ, UK}

\maketitle

\section{Nonrephasing beat maps}

\hspace{1em}

\begin{figure}[h]
    \centering
    \includegraphics{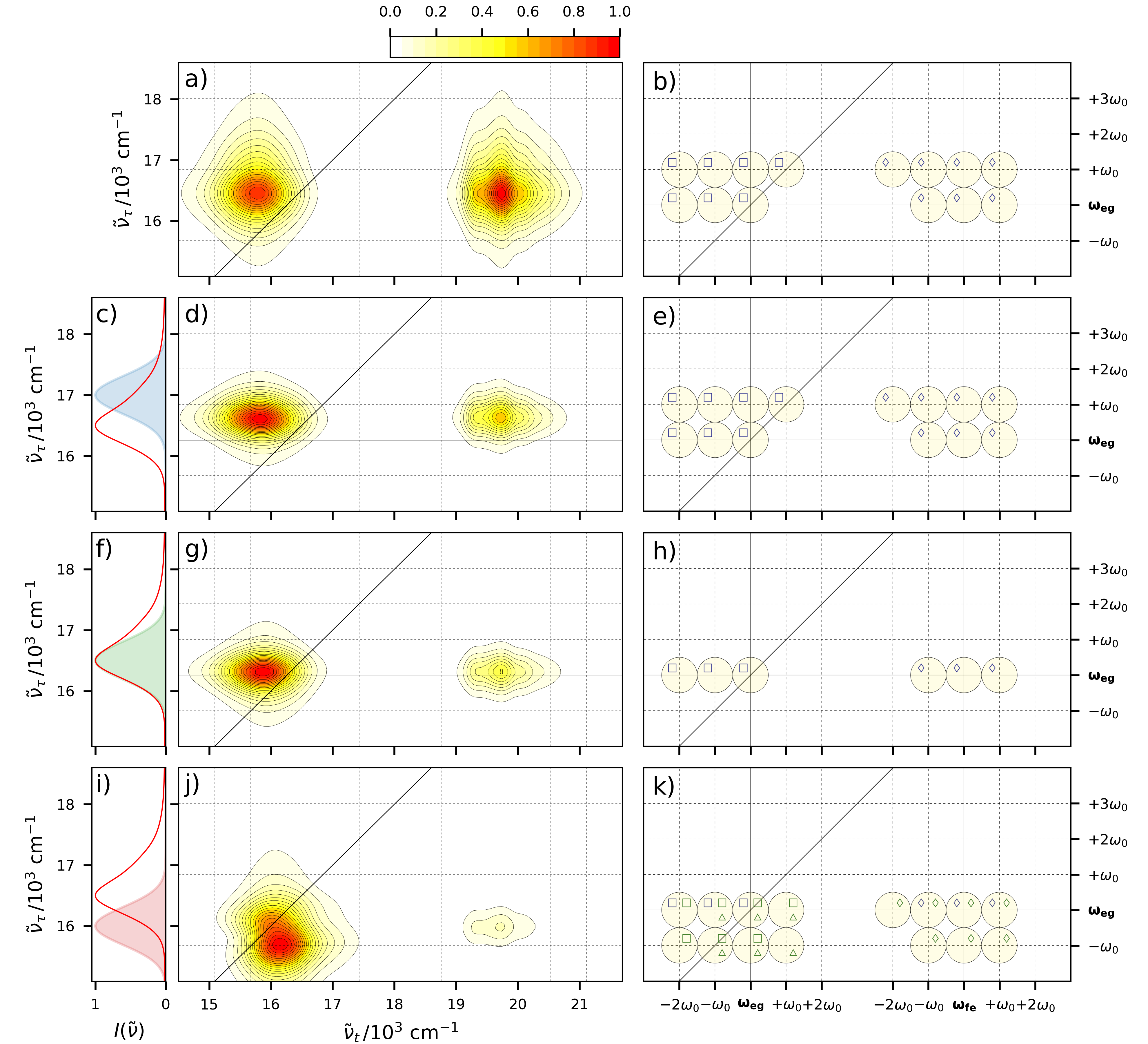}
    \caption{(a, d, g, j) Nonrephasing $+\SI{585}{\wn}$ beat maps and (b, e, h, k) corresponding location key diagrams for the impulsive limit, blue shifted, centre and red shifted pump spectra, respectively. (c) Blue shifted, (f) centre and (i) red shifted pump spectra (shaded) overlaid with the steady state absorption spectrum (red line).}
    \label{fig:pump_NonrephPstv}
\end{figure}

\begin{figure}[h]
    \centering
    \includegraphics{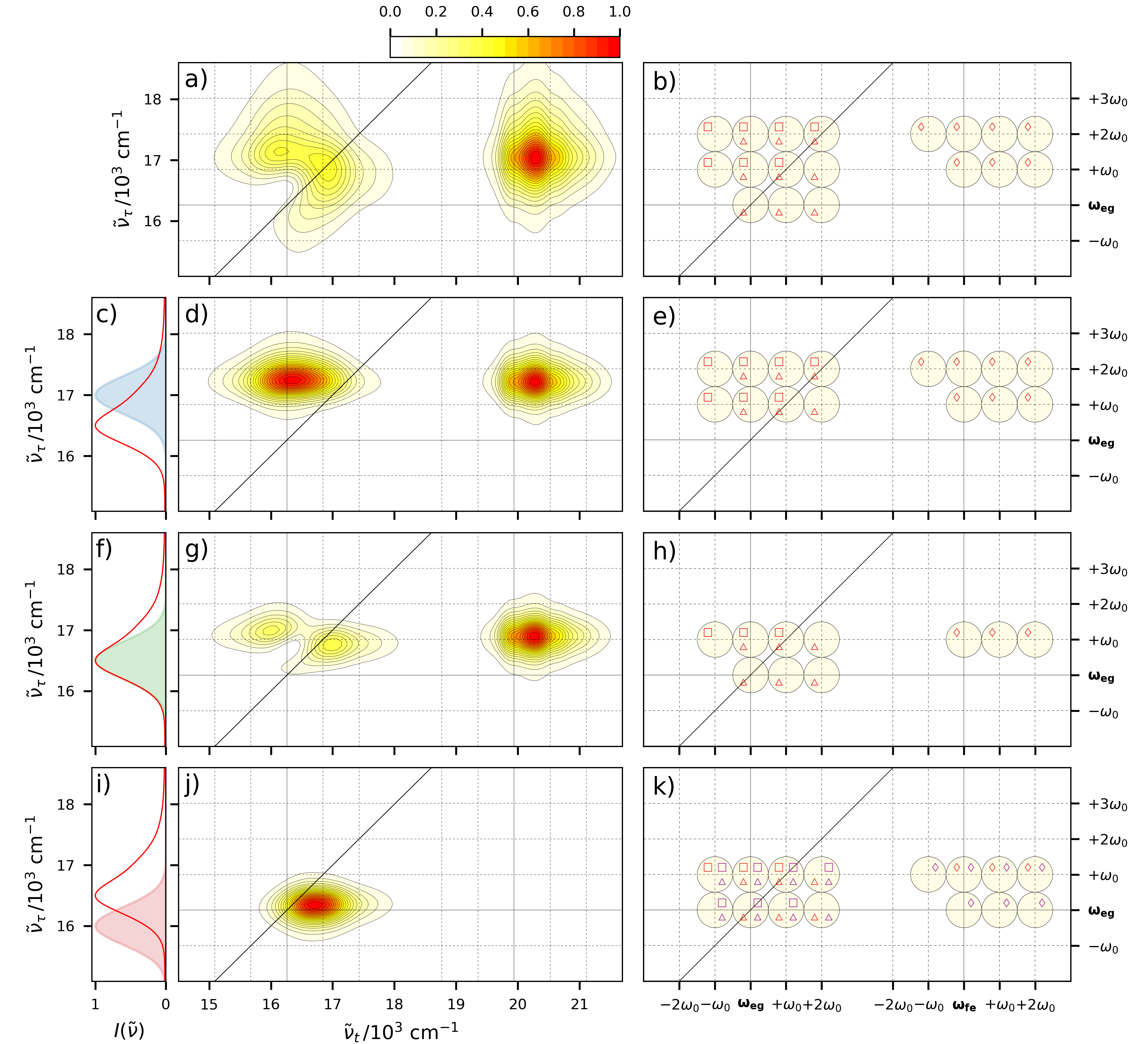}
    \caption{(a, d, g, j) Nonrephasing $-\SI{585}{\wn}$ beat maps and (b, e, h, k) corresponding location key diagrams for the impulsive limit, blue shifted, centre and red shifted pump spectra, respectively. (c) Blue shifted, (f) centre and (i) red shifted pump spectra (shaded) overlaid with the steady state absorption spectrum (red line).}
    \label{fig:pump_NonrephNgtv}
\end{figure}

\section{Double-sided Feynman diagrams}

\renewcommand{\arraystretch}{1.5}
    \begin{table}[ht!]
        \centering
        \begin{tabular}{c|c|c|c}
            & GSB & SE & ESA \\
            \hline
            $\color{Blue}+\omega_x\color{black}$ starting $\ket{g0}\!\bra{g0}$ & \protect\marksymbol{triangle}{Blue} & \protect\marksymbol{square}{Blue} & \protect\marksymbol{diamond}{Blue} \\
            $\color{Red}-\omega_x\color{black}$ starting $\ket{g0}\!\bra{g0}$ & \protect\marksymbol{triangle}{Red} & \protect\marksymbol{square}{Red} & \protect\marksymbol{diamond}{Red} \\
            $\color{OliveGreen}+\omega_x\color{black}$ starting $\ket{g1}\!\bra{g1}$ & \protect\marksymbol{triangle}{OliveGreen} & \protect\marksymbol{square}{OliveGreen} & \protect\marksymbol{diamond}{OliveGreen} \\
            $\color{Mulberry}-\omega_x\color{black}$ starting $\ket{g1}\!\bra{g1}$ & \protect\marksymbol{triangle}{Mulberry} & \protect\marksymbol{square}{Mulberry} & \protect\marksymbol{diamond}{Mulberry}
        \end{tabular}
        \caption{Symbol key for Liouville pathways.}
        \label{SItab:PathwayKey}
    \end{table}

\vspace{2em} 

\begin{figure}[h]
    \centering
    \includegraphics[width=\textwidth]{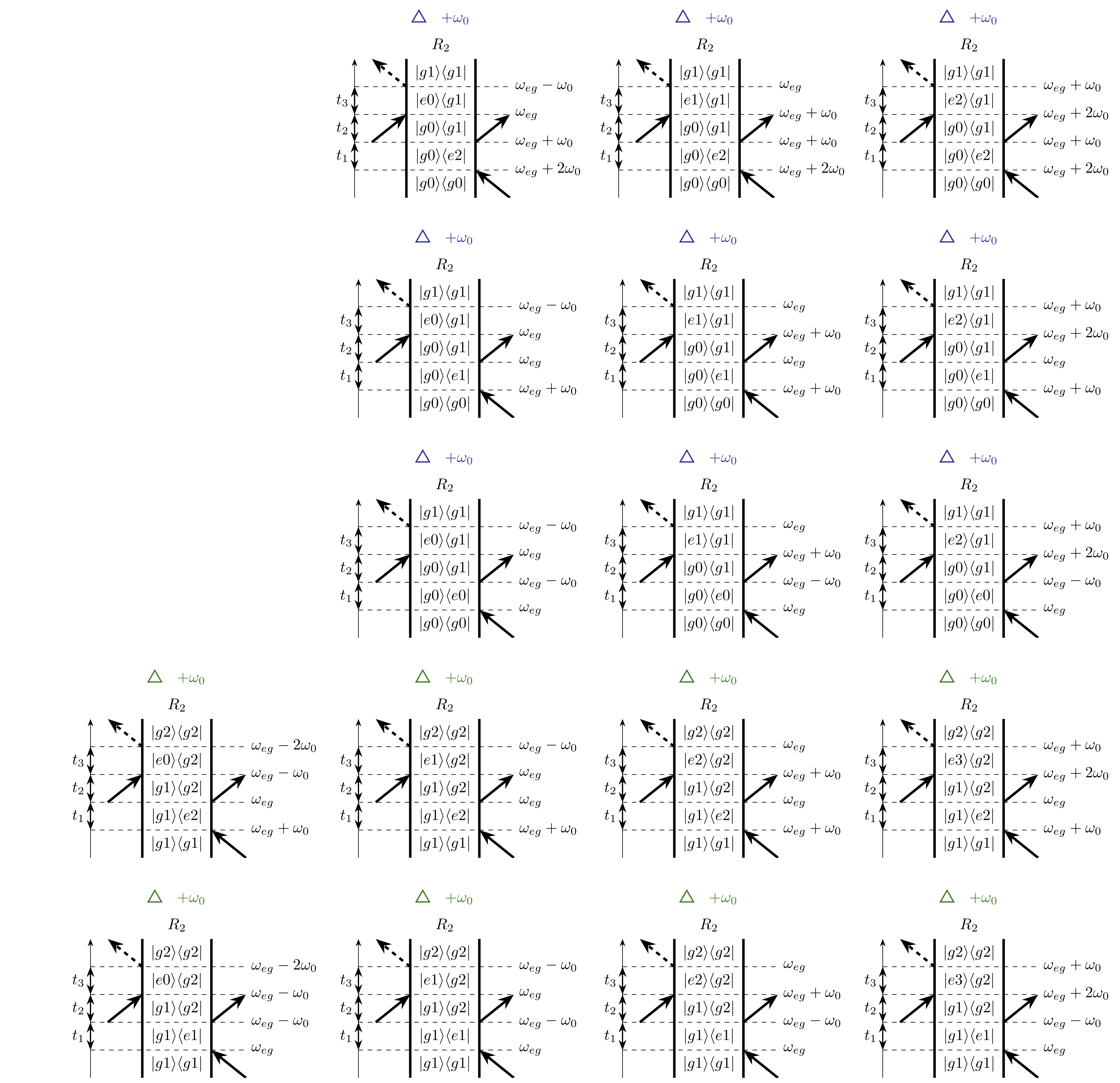}
    \caption{Rephasing $+\omega_0$ GSB pathways labelled as per table~\ref{SItab:PathwayKey}.}
    \label{fig:R_+585_GSB}
\end{figure}

\begin{figure}[h]
    \centering
    \includegraphics[width=\textwidth]{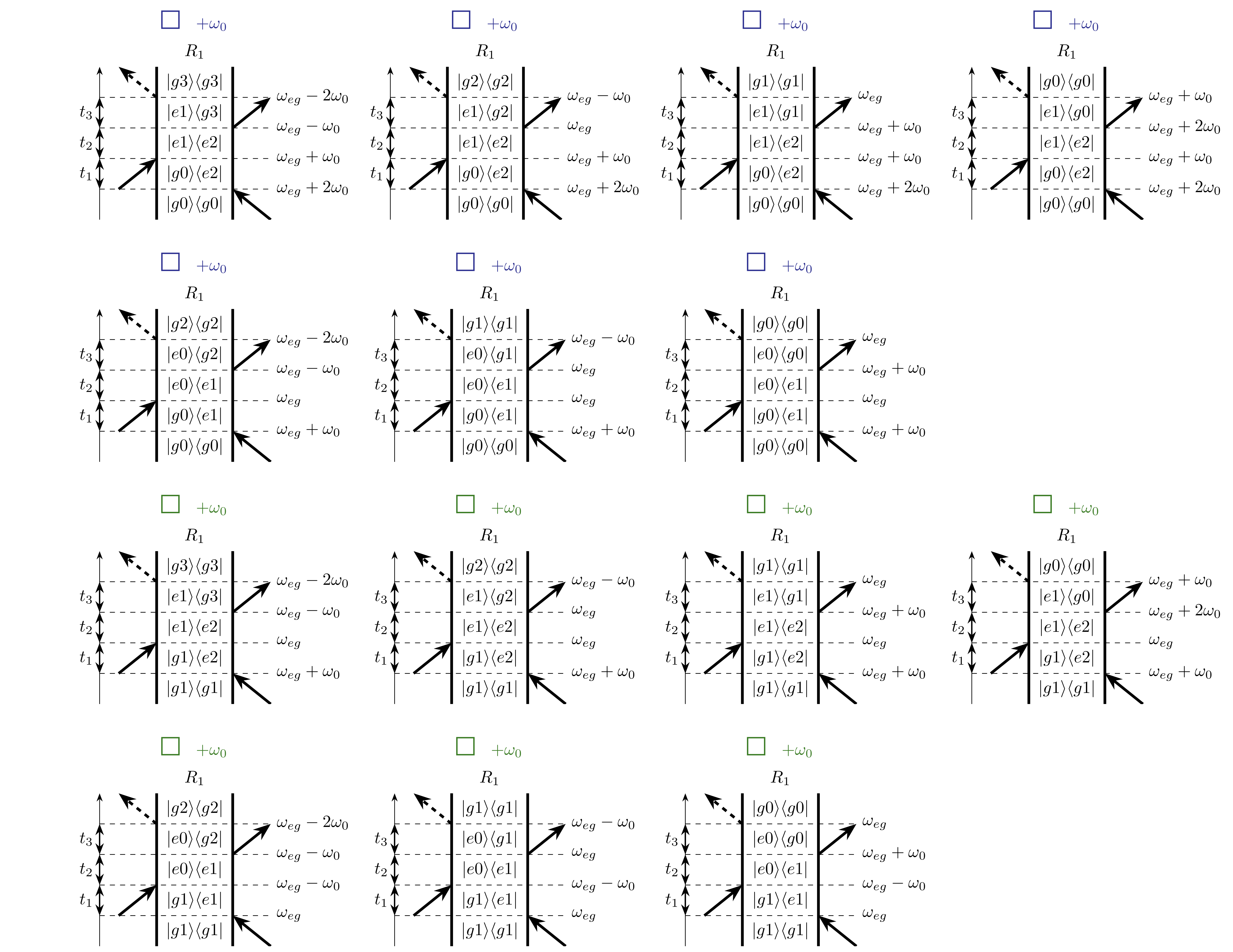}
    \caption{Rephasing $+\omega_0$ SE pathways labelled as per table~\ref{SItab:PathwayKey}.}
    \label{fig:R_+585_SE}
\end{figure}

\begin{figure}[h]
    \centering
    \includegraphics[width=\textwidth]{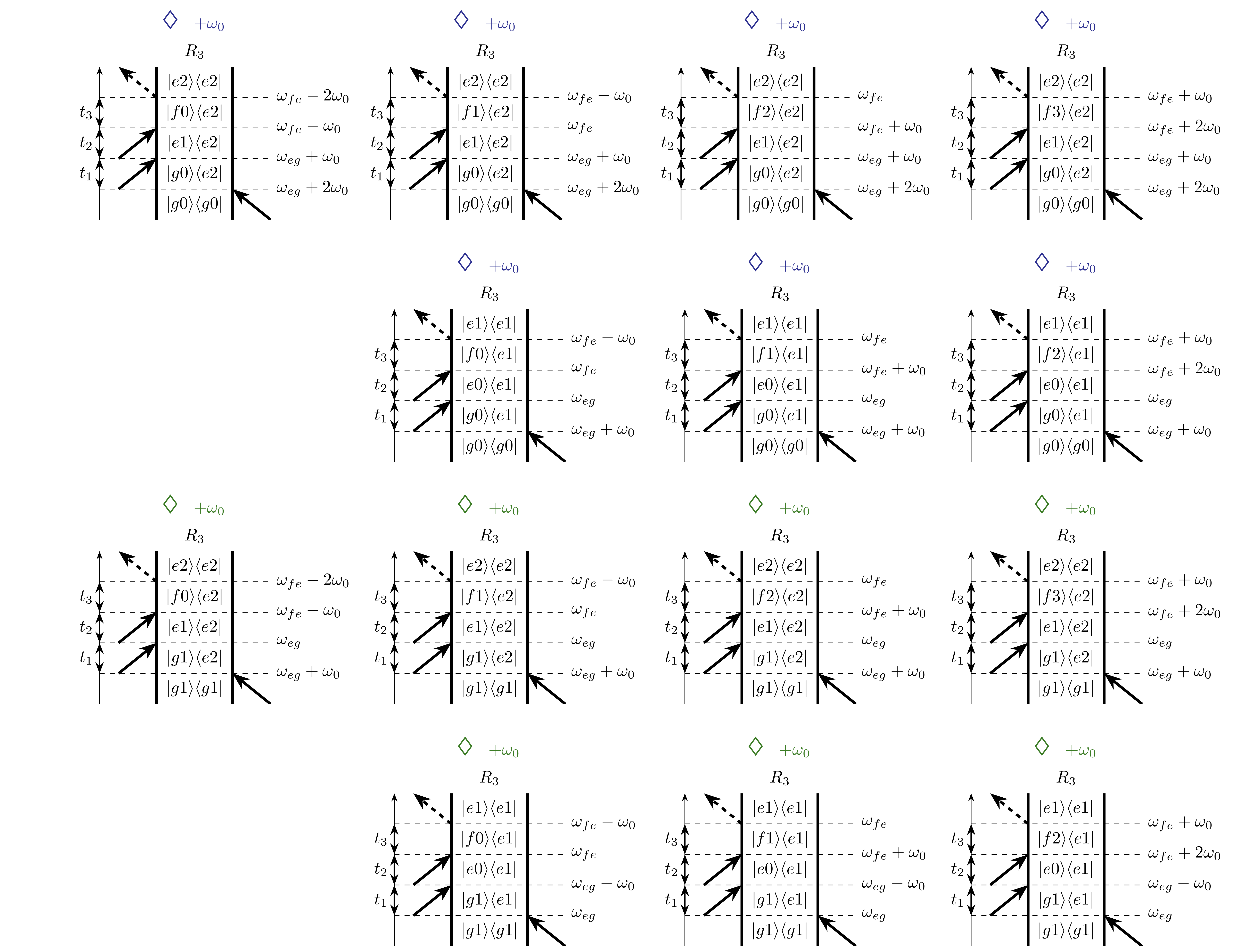}
    \caption{Rephasing $+\omega_0$ ESA pathways labelled as per table~\ref{SItab:PathwayKey}.}
    \label{fig:R_+585_ESA}
\end{figure}

\begin{figure}[h]
    \centering
    \includegraphics[width=0.75\textwidth]{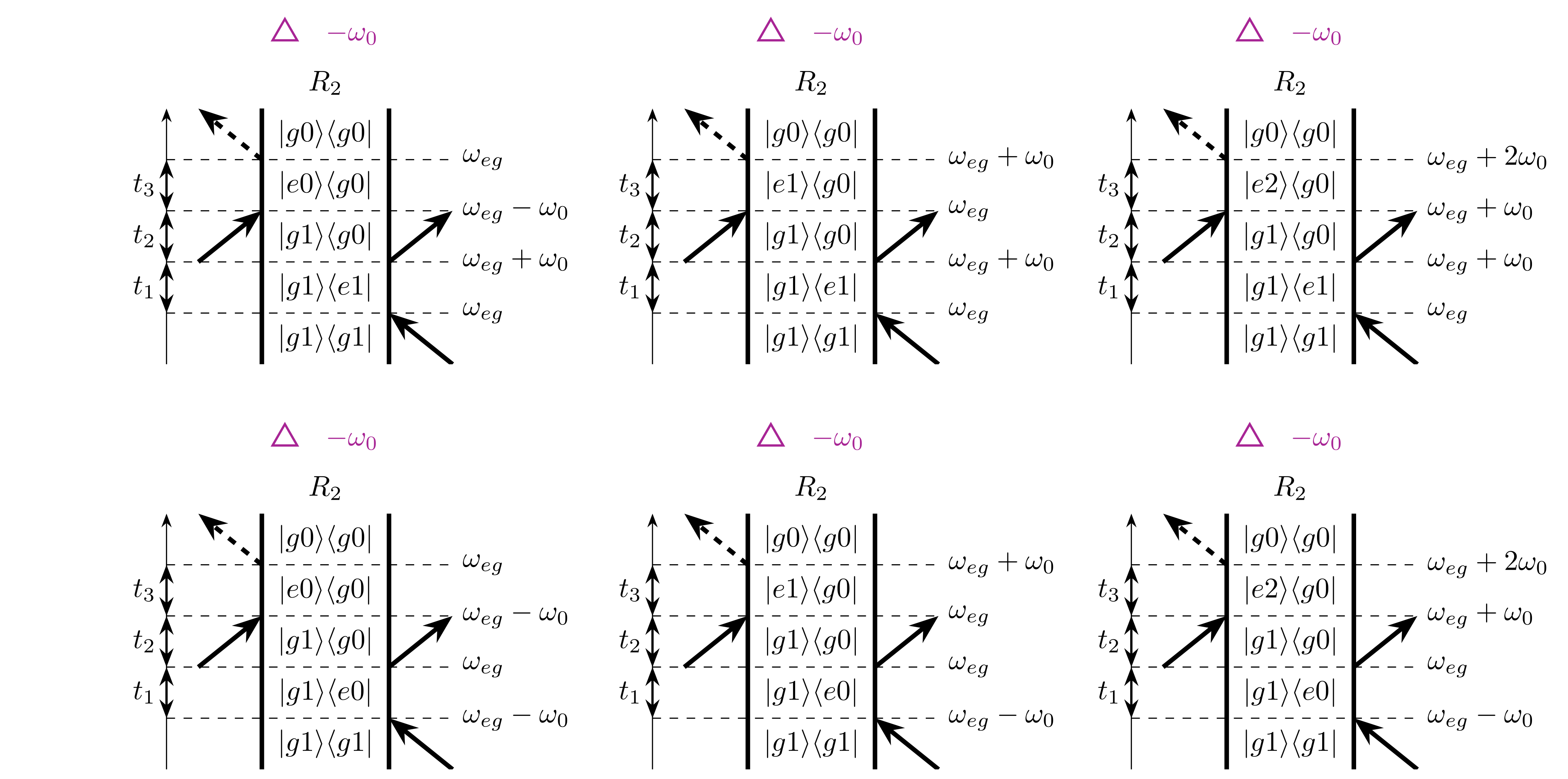}
    \caption{Rephasing $-\omega_0$ GSB pathways labelled as per table~\ref{SItab:PathwayKey}.}
    \label{fig:R_-585_GSB}
\end{figure}

\begin{figure}[h]
    \centering
    \includegraphics[width=\textwidth]{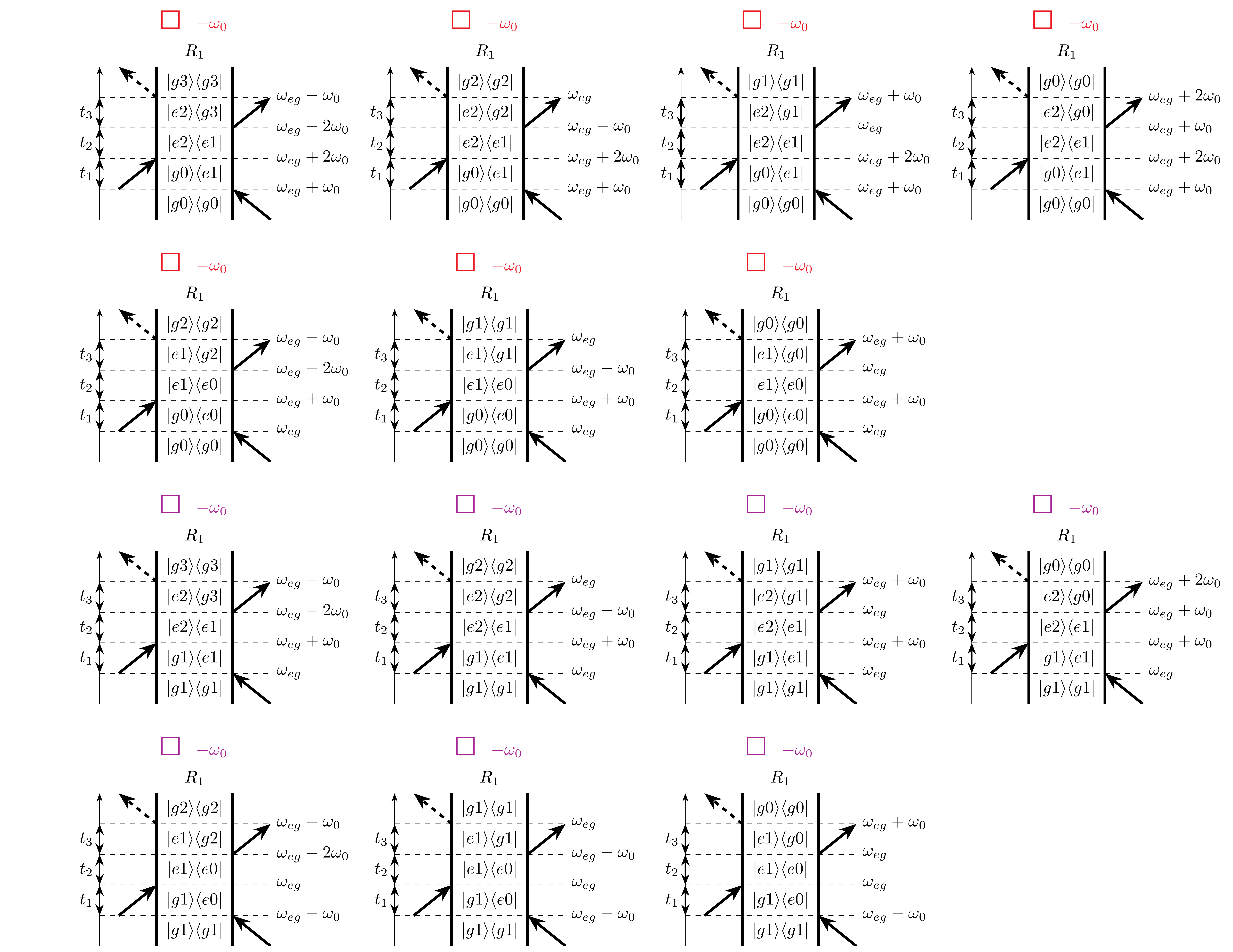}
    \caption{Rephasing $-\omega_0$ SE pathways labelled as per table~\ref{SItab:PathwayKey}.}
    \label{fig:R_-585_SE}
\end{figure}

\begin{figure}[h]
    \centering
    \includegraphics[width=\textwidth]{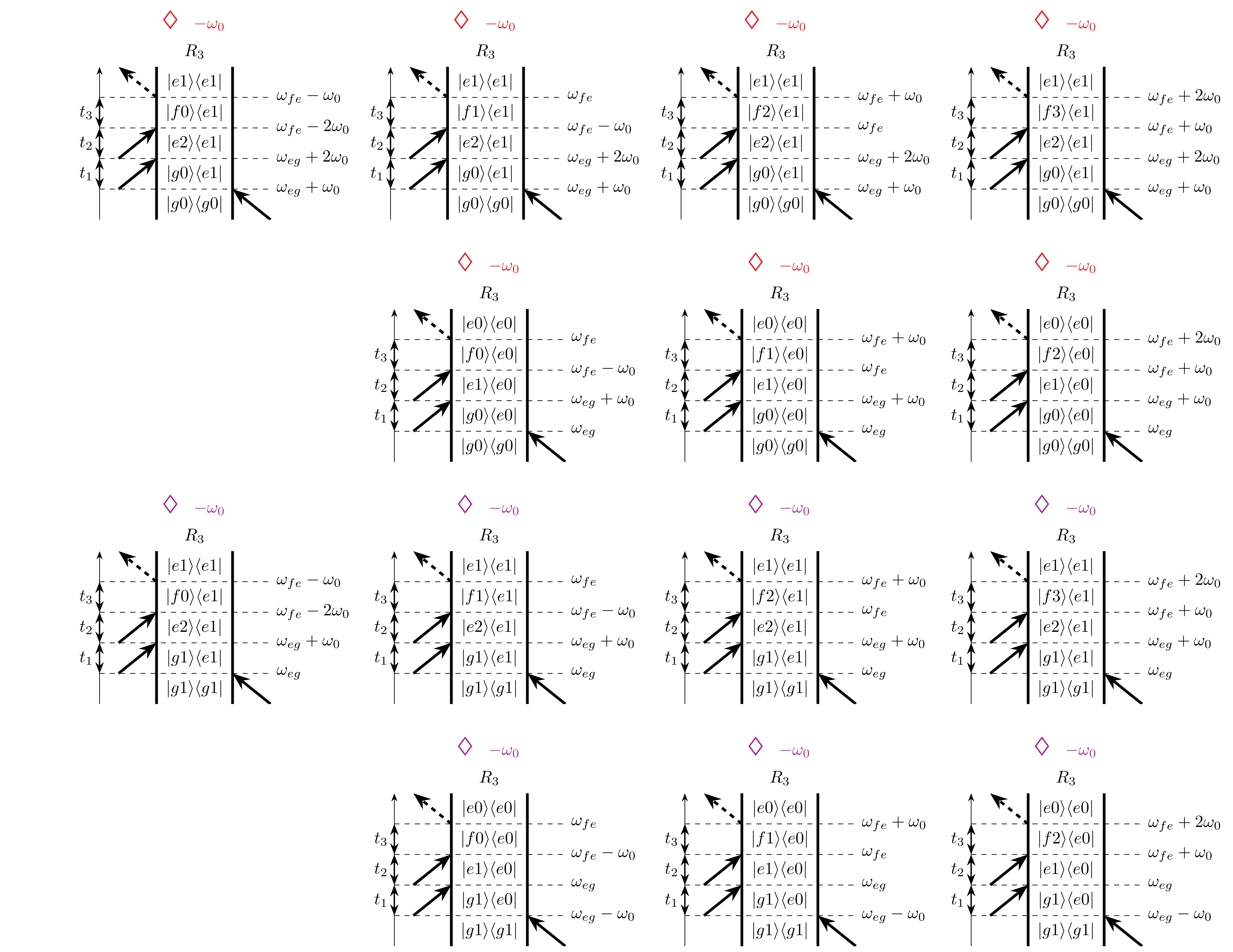}
    \caption{Rephasing $-\omega_0$ ESA pathways labelled as per table~\ref{SItab:PathwayKey}.}
    \label{fig:R_-585_ESA}
\end{figure}

\begin{figure}[h]
    \centering
    \includegraphics[width=0.75\textwidth]{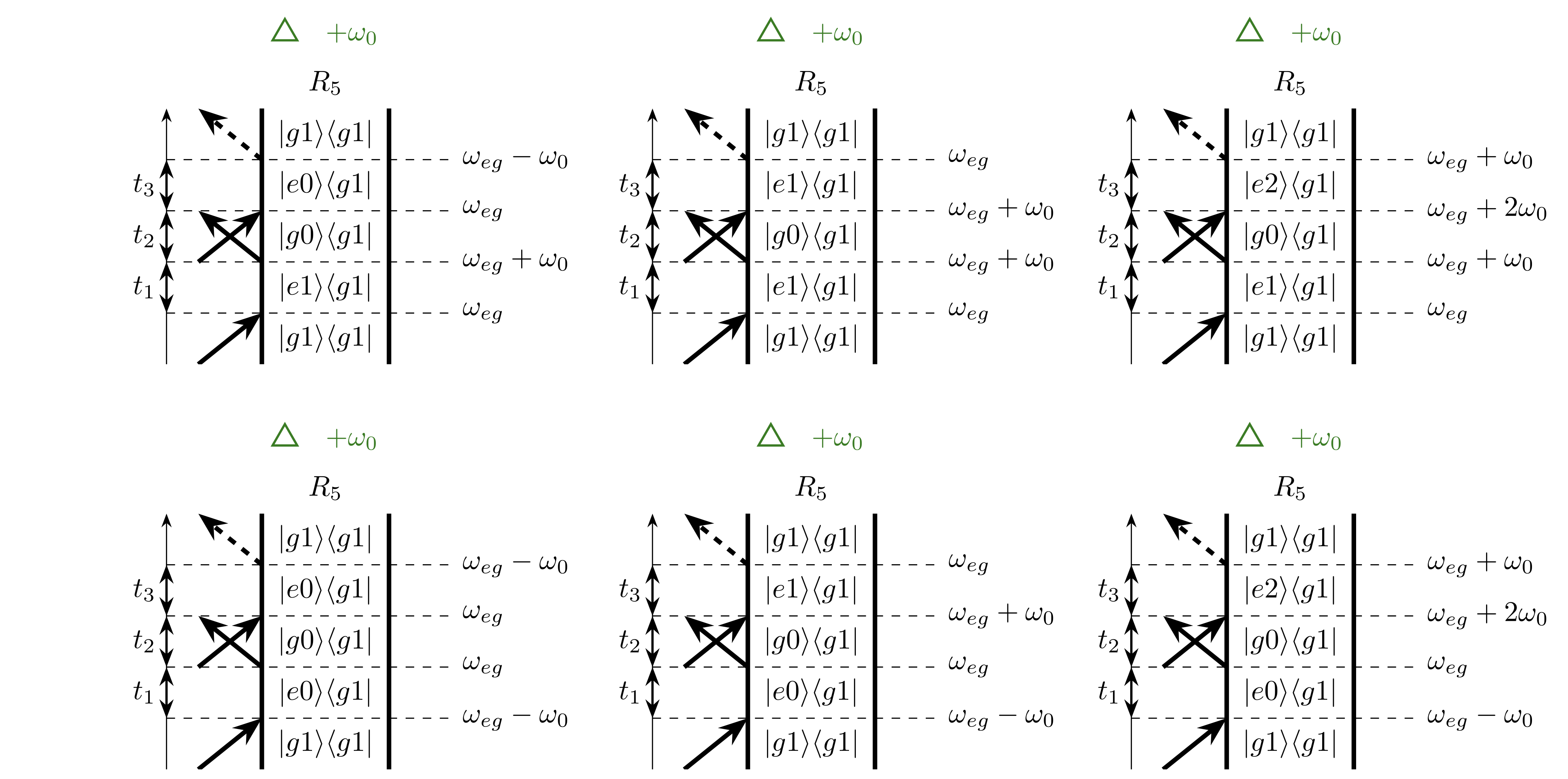}
    \caption{Nonrephasing $+\omega_0$ GSB pathways labelled as per table~\ref{SItab:PathwayKey}.}
    \label{fig:NR_+585_GSB}
\end{figure}

\begin{figure}[h]
    \centering
    \includegraphics[width=\textwidth]{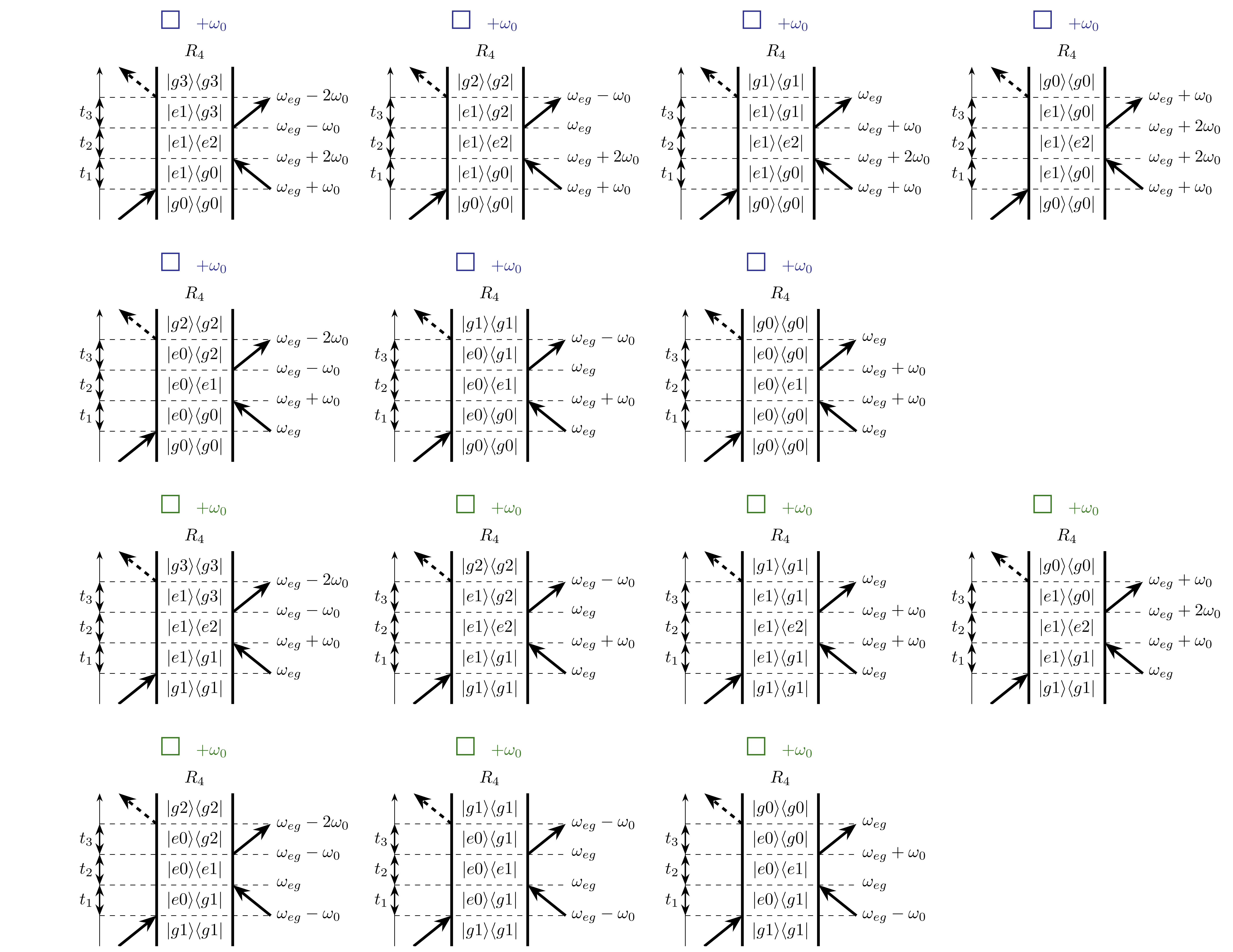}
    \caption{Nonrephasing $+\omega_0$ SE pathways labelled as per table~\ref{SItab:PathwayKey}.}
    \label{fig:NR_+585_SE}
\end{figure}

\begin{figure}[h]
    \centering
    \includegraphics[width=\textwidth]{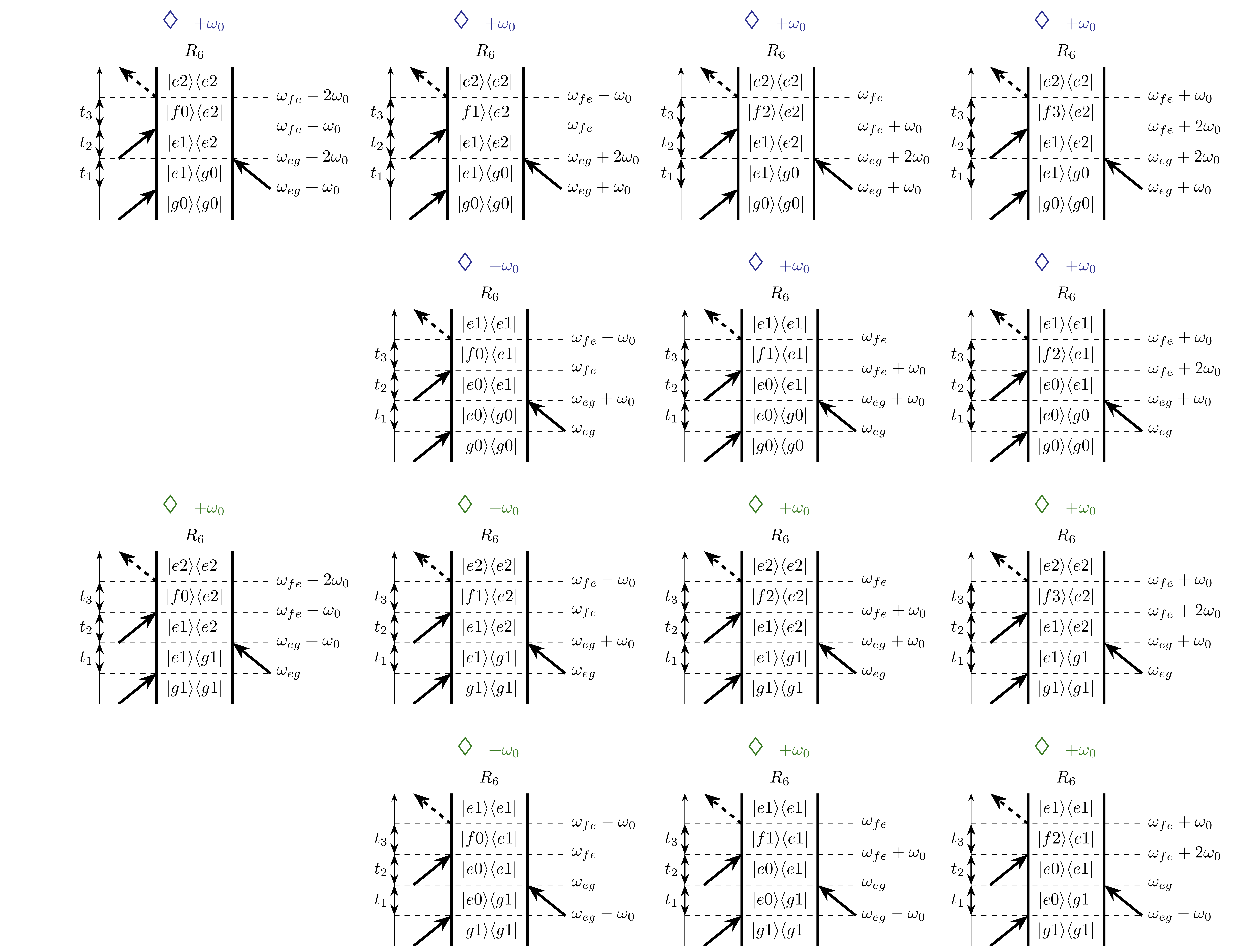}
    \caption{Nonrephasing $+\omega_0$ ESA pathways labelled as per table~\ref{SItab:PathwayKey}.}
    \label{fig:NR_+585_ESA}
\end{figure}

\begin{figure}[h]
    \centering
    \includegraphics[width=\textwidth]{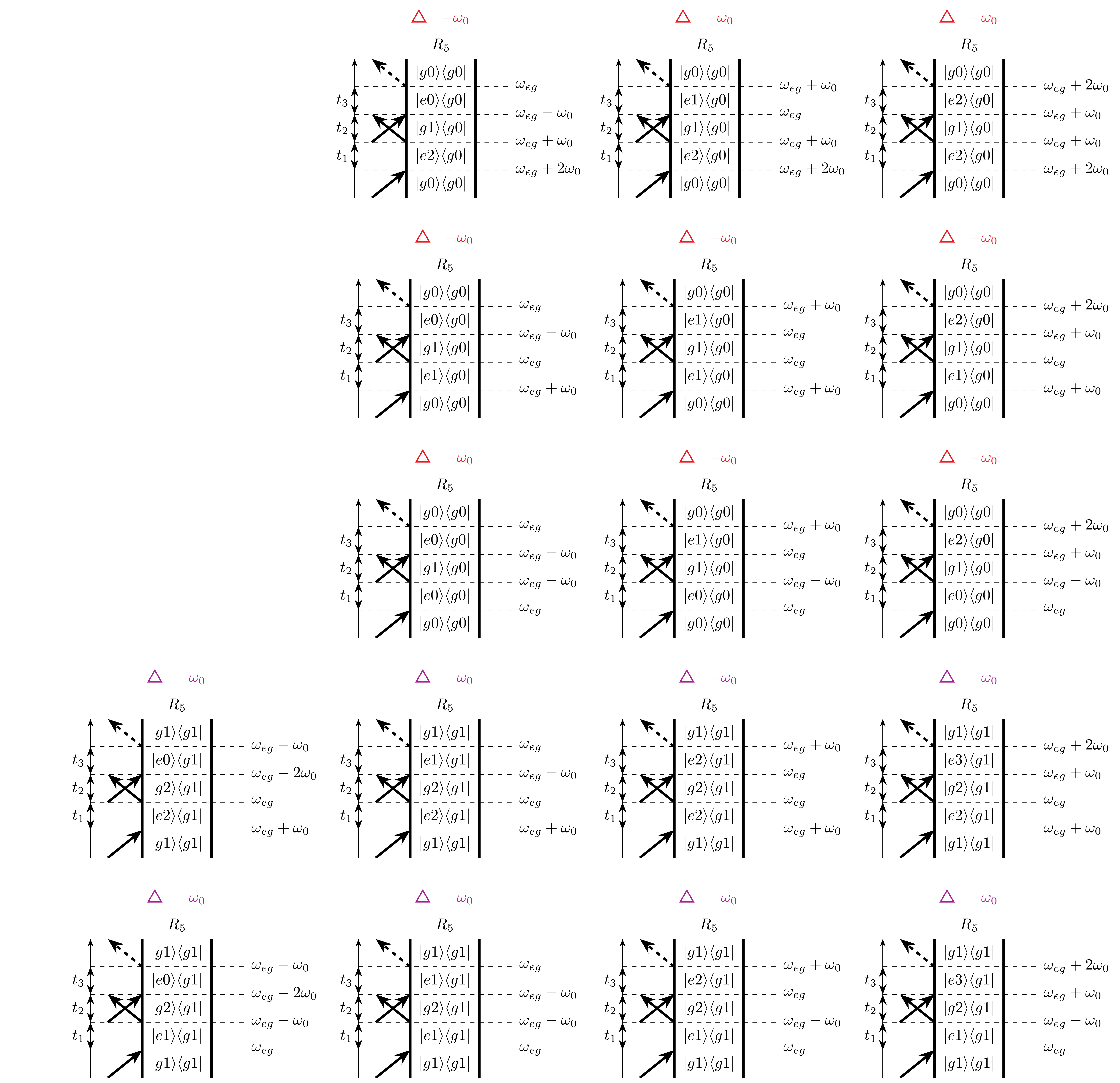}
    \caption{Nonrephasing $-\omega_0$ GSB pathways labelled as per table~\ref{SItab:PathwayKey}.}
    \label{fig:NR_-585_GSB}
\end{figure}

\begin{figure}[h]
    \centering
    \includegraphics[width=\textwidth]{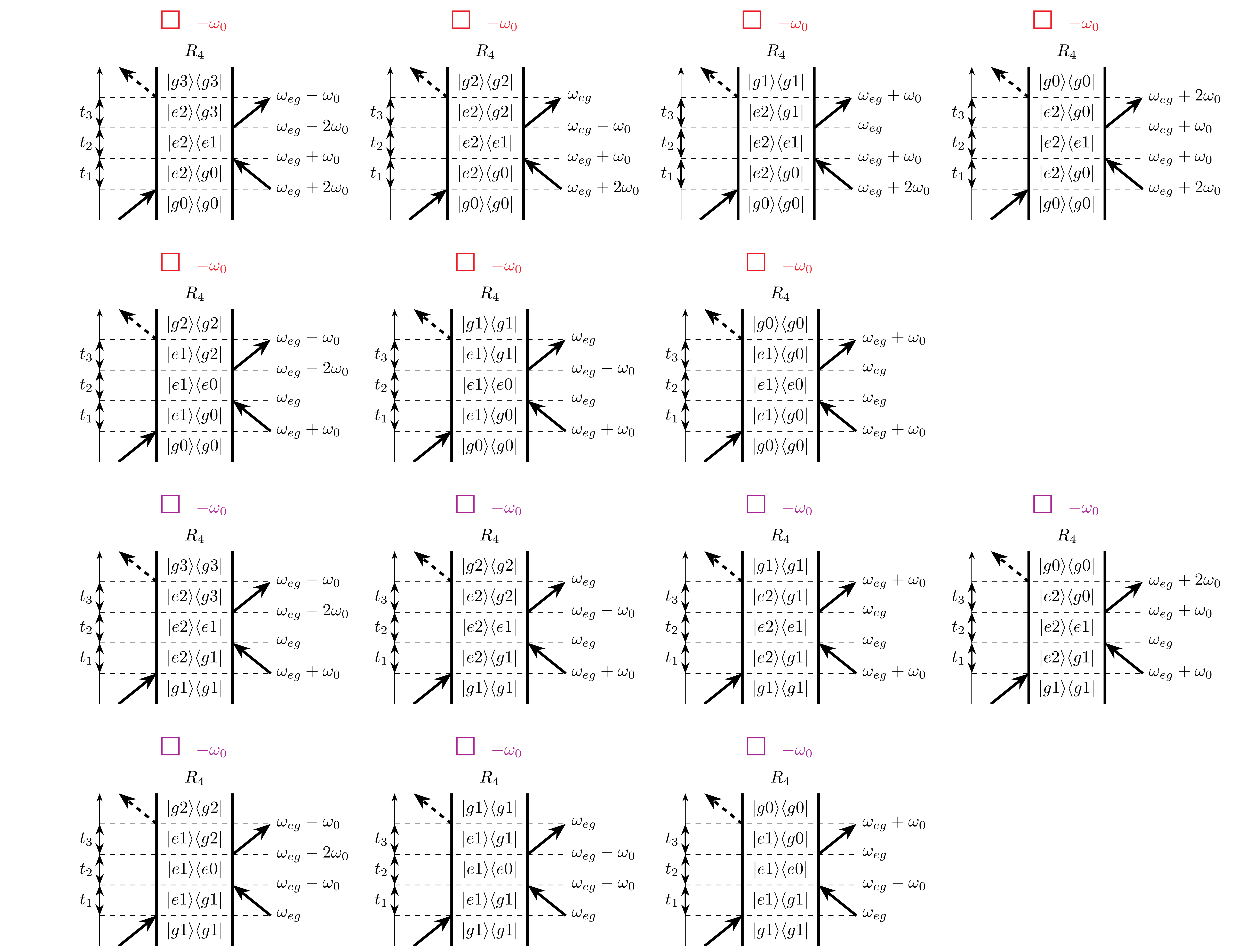}
    \caption{Nonrephasing $-\omega_0$ SE pathways labelled as per table~\ref{SItab:PathwayKey}.}
    \label{fig:NR_-585_SE}
\end{figure}

\begin{figure}[h]
    \centering
    \includegraphics[width=\textwidth]{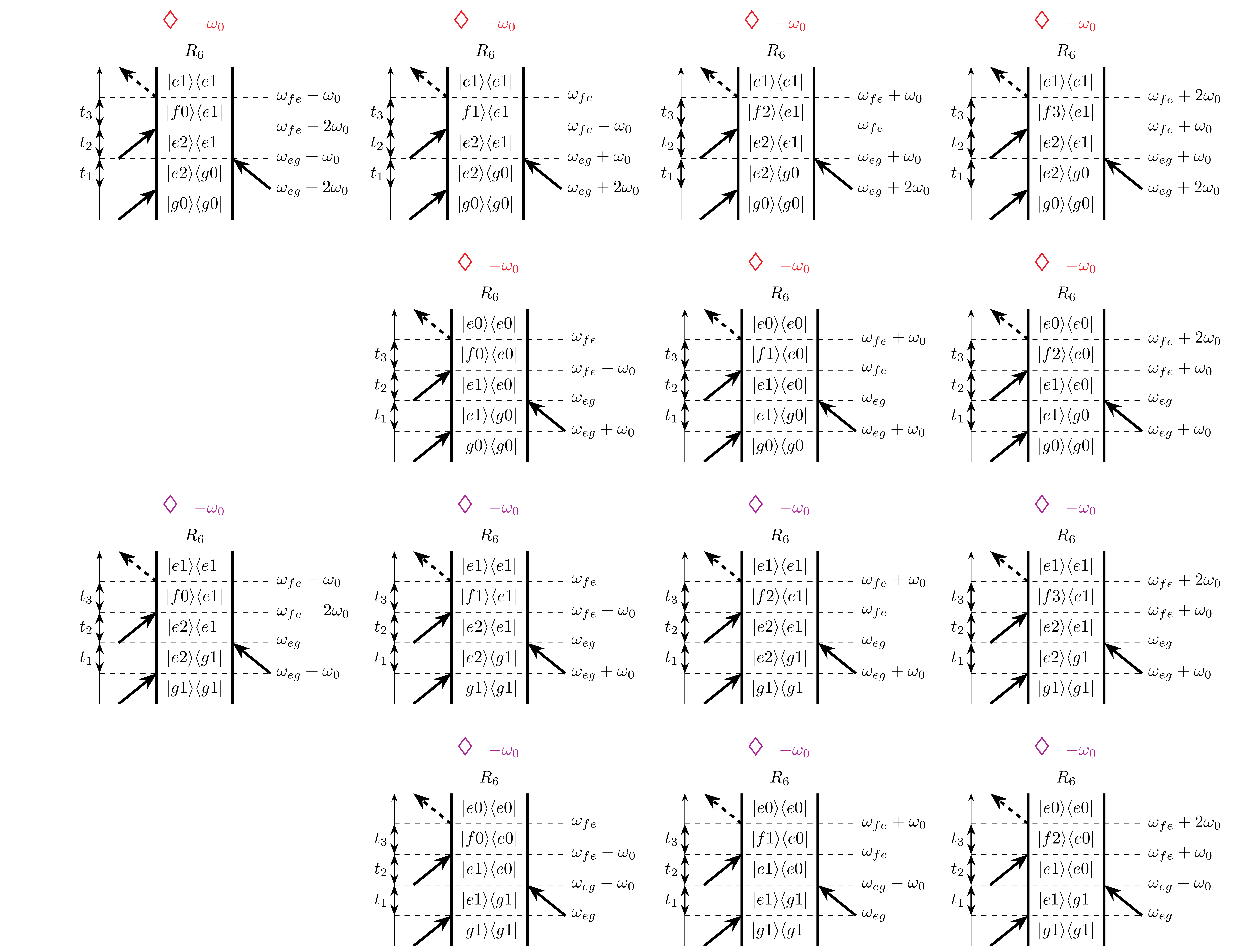}
    \caption{Nonrephasing $-\omega_0$ ESA pathways labelled as per table~\ref{SItab:PathwayKey}.}
    \label{fig:NR_-585_ESA}
\end{figure}

    Double-sided Feynman diagrams for all the pathways identified in figures 6, 7, \ref{fig:pump_NonrephPstv} and \ref{fig:pump_NonrephNgtv} are presented in figures \ref{fig:R_+585_GSB} -- \ref{fig:NR_-585_ESA}. Ground state bleach (GSB), stimulated emission (SE) and excited state absorption (ESA) pathways are labelled with colour-coded symbols defined in table \ref{SItab:PathwayKey}, for which the corresponding rephasing (R) and nonrephasing (NR) molecular response functions are,

    \begin{equation}
        R^{(3)}_\text{R, GSB}(\tau,T,t)=\left(\frac{i}{\hbar}\right)^3\text{Tr}\left(\hat\mu G(t+T+\tau,T+\tau)\hat\mu_+ G(T+\tau,\tau) (G(\tau,0)\rho_\mathbf{j}(-\infty)\hat\mu_-)\hat\mu_+\right),
    \end{equation}
    
    \begin{equation}
        R^{(3)}_\text{R, SE}(\tau,T,t)=\left(\frac{i}{\hbar}\right)^3\text{Tr}\left(\hat\mu G(t+T+\tau,T+\tau) (G(T+\tau,\tau)\hat\mu_+ (G(\tau,0)\rho_\mathbf{j}(-\infty)\hat\mu_-))\hat\mu_+\right),
    \end{equation}
    
    \begin{equation}
        R^{(3)}_\text{R, ESA}(\tau,T,t)=\left(\frac{i}{\hbar}\right)^3\text{Tr}\left(\hat\mu G(t+T+\tau,T+\tau)\hat\mu_+ G(T+\tau,\tau)\hat\mu_+ G(\tau,0)\rho_\mathbf{j}(-\infty)\hat\mu_-\right),
    \end{equation}

    \begin{equation}
        R^{(3)}_\text{NR, GSB}(\tau,T,t)=\left(\frac{i}{\hbar}\right)^3\text{Tr}\left(\hat\mu G(t+T+\tau,T+\tau)\hat\mu_+ G(T+\tau,\tau)\hat\mu_- G(\tau,0)\hat\mu_+\rho_\mathbf{j}(-\infty)\right),
    \end{equation}

    \begin{equation}
        R^{(3)}_\text{NR, SE}(\tau,T,t)=\left(\frac{i}{\hbar}\right)^3\text{Tr}\left(\hat\mu G(t+T+\tau,T+\tau) (G(T+\tau,\tau) (G(\tau,0)\hat\mu_+\rho_\mathbf{j}(-\infty))\hat\mu_-)\hat\mu_+\right),
    \end{equation}

    \begin{equation}
        R^{(3)}_\text{NR, ESA}(\tau,T,t)=\left(\frac{i}{\hbar}\right)^3\text{Tr}\left(\hat\mu G(t+T+\tau,T+\tau)\hat\mu_+ G(T+\tau,\tau) (G(\tau,0)\hat\mu_+\rho_\mathbf{j}(-\infty))\hat\mu_-\right).
    \end{equation}

\bibliography{references}